\documentclass[10pt, reprint, twocolumn, amssymb, amsmath, nofootinbib, tightenlines, floatfix, longbibliography, citeautoscript, amsmath, groupedaddress, ajp]{revtex4-1}

\usepackage{fancyhdr}
\fancyheadoffset{0cm}
\usepackage{comment}
\usepackage{bm}
\usepackage{amsmath}
\usepackage{amsfonts}
\usepackage{amssymb}
\usepackage{float}
\usepackage{graphicx}
\usepackage{dcolumn}
\usepackage{mathtools}

\usepackage{hyperref}
\hypersetup{
    colorlinks=true,
    linktocpage=true,
    linkcolor=blue,
    filecolor=magenta,      
    urlcolor=blue,
    citecolor=blue,
    bookmarks=true,
}

\makeatletter
\renewcommand*\l@section{\@dottedtocline{1}{1.5em}{2.3em}}
\makeatother

\graphicspath{{Graphics/}}

\newcommand{\be}{\begin}
\newcommand{\en}{\end}
\newcommand{\eq}{equation}

\begin{document}

\lhead{\small{D. M. Riffe and J. D. Christensen}}
\rhead{\small{Alkali Metal Vibrational Dynamics}}
\cfoot{--\thepage--}

\title{Alkali-metal vibrations in bcc, fcc, hcp, and 9R structures:  Implications for the energetics of Li and Na martensitic phases}

\begin{abstract}
We present an embedded-atom-method (EAM) model that is specifically designed to accurately describe vibrations in bcc alkali metals.  Using this model, we study bulk vibrational structure of Li, Na, K, and Rb when configured in bcc and the closed-packed (cp) fcc, hcp, and 9$R$ phases.  From the vibrational density of states for each phase we thence find the corresponding vibrational contribution $A_{\rm vib}(T)$ to the Helmholtz free energy $A(T)$.  Utilizing (i) differences in $A_{\rm vib}(T)$ between the bcc and cp structures and (ii) experimentally inferred thermodynamic transition temperatures for Li and Na (which martensitically transform from bcc to cp phases upon cooling), we extract values for zero-temperature energy differences between the bcc and relevant cp phases.  We also put constraints on zero-temperature cp-bcc energy differences for K and Rb, which do not exhibit temperature induced transitions.  
\end{abstract}

\author{D. M. Riffe}
\email[Author to whom correspondence should be addressed; electronic mail: ]{mark.riffe@usu.edu}

\author{Jake D. Christensen}

\affiliation{Physics Department, Utah State University, Logan, UT 84322, USA}

\date{\today}

\maketitle

\section{Introduction}
\label{SecI}

In 1947 C.~S.~Barrett reported a phase transition in Li upon cooling:  in the vicinity of 77 K a Li sample---when plastically deformed---martensitically transforms from its room-temperature (RT) bcc structure to an fcc structure \cite{barrett1947}.  Upon subsequent heating to $\sim$156 K, the sample transformed back to the bcc structure.  Key elements of Barrett's initial report are (i) the transformation between bcc and a closed-packed (cp) structure and (ii) hysteresis associated with the transformation.  Barrett also offered a salient suggestion for the energetics of the transformation:  at higher temperatures the Helmholtz free energy $A(T) \!= \!U \! - \! TS$ of the bcc phase is lower than that of the fcc phase owing to larger entropy $S$ arising from vibrational modes associated with the elastic constant $C' \!= \! (C_{11} \! - C_{12}) / 2$, which is often much smaller in the bcc form of materials that exhibit both bcc and fcc phases.

\begin{figure*}[t]
\centerline{\includegraphics[scale=1]{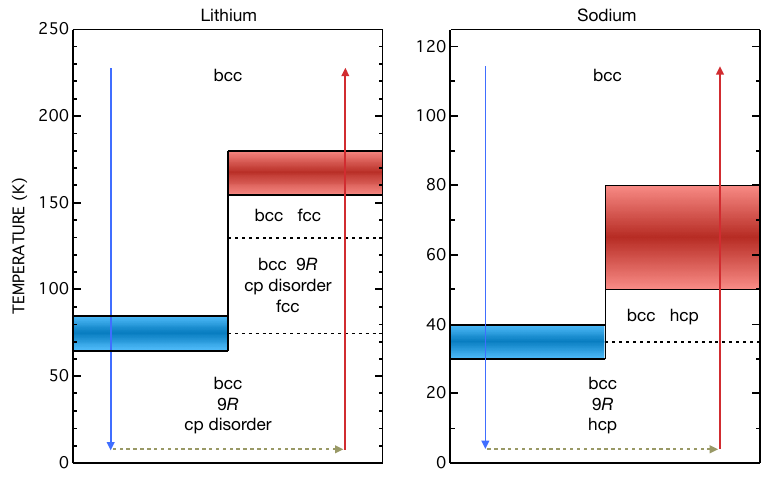}}
\caption{Graphical representation of Li and Na phases and transitions between phases upon cooling (blue downward arrow) and subsequent heating (red upward arrow).  Dashed lines indicate approximate temperatures for structural changes.  Experimentally observed regions of transition from pure bcc (upon cooling) and back to pure bcc (upon heating) are represented by the blue and red shaded bands, respectively. }
\label{Fig1a}
\end{figure*}

This initial report by Barrett inspired a number of subsequent structural investigations of the alkali metals.  It was discovered that Li \cite{barrett1948,hovi1966,mccarthy1980,berliner1986A,smith1987,schwarz1990,schwarz1991,berliner1986B,berliner1989,smith1994,maier1995,ackland2017,owen1954,ernst1986,BarrettAC1956} and Na \cite{BarrettAC1956,stedman1976,schwarz1992,berliner1992,abe1994,abe1997,blaschko1984,szente1988} undergo martensitic phase transitions when sufficiently cooled, even without a perturbation such as plastic deformation.  Several cp structures are manifest upon cooling and subsequent reheating, as illustrated in Fig.~\ref{Fig1a}.  For Li the dominant cp phases are fcc, 9$R$ (which comprises the cp-plane stacking sequence $ABCBCACAB$), and cp disordered polytypes \cite{overhauser1984,berliner1986A,berliner1986B,smith1987,berliner1989,schwarz1990,schwarz1991}, while for Na the dominant cp phases are hcp and 9$R$ \cite{BarrettAC1956,stedman1976,schwarz1992,abe1994,berliner1992}.  In contrast, the heavier alkali metals K, Rb, and Cs maintain bcc structure upon cooling \cite{BarrettAC1956,berliner1989,kubinski1993}.

The observation of multiple cp phases in Li and Na prompted theoretical investigation of differences in $A(T)$ between the bcc and various cp structures, with the aim of predicting thermodynamic transition temperatures $T_{\rm bcc}^{\rm \, cp}$ between the bcc and  relevant cp phases.  Unfortunately, no theoretical consensus has emerged from these calculations.  For Li, calculated values of $T_{\rm bcc}^{\rm \, fcc}$ range from 60  to 217 K \cite{schneider1970,bajpai1975,ackland2017,hutcheon2019}.  One study even predicts that no transition exists between bcc and fcc phases \cite{staikov1997}.  For the bcc to 9$R$ transition, calculated values of $T_{\rm bcc}^{\, 9R}$ for Li vary from 68 to 220 K \cite{yao2009,Liu1999,ackland2017,hutcheon2019}.  Similar variations exist for Na:  $T_{\rm bcc}^{\rm \, hcp}$ values equal to 99 K \cite{pynn1971} and  260 K \cite{straub1971} have been reported.  For K, $T_{\rm bcc}^{\rm \, fcc}$ and $T_{\rm bcc}^{\rm \, hcp}$ values of 70 K and 92 K, respectively, have been calculated \cite{pynn1971}, even though no close-packed structures are observed for this metal. 

\begin{figure}[b]
\centerline{\includegraphics[scale=0.6]{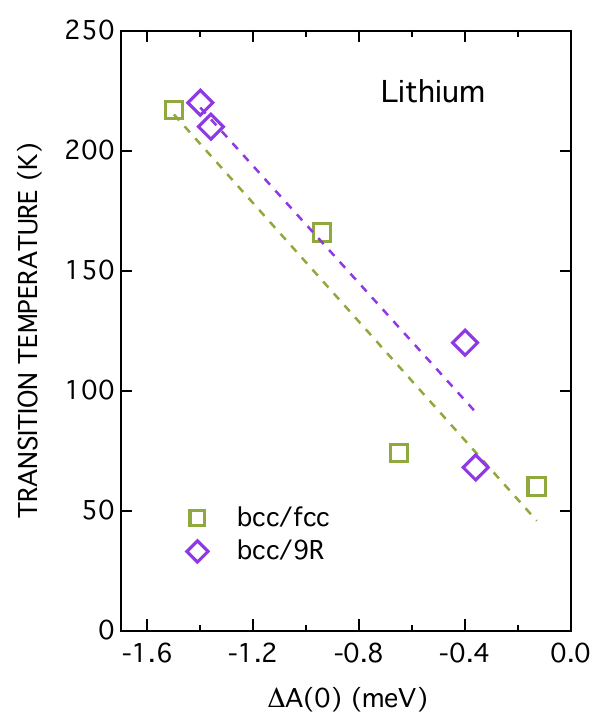}}
\caption{Values of thermodynamic transition temperatures $T_{\rm bcc}^{\rm \, fcc}$ and $T_{\rm bcc}^{\, 9R}$ vs zero-temperature Helmholtz-energy differences $\Delta A(0)$ for Li.  All data are from first-principle-theory (FPT) calculations \cite{schneider1970,bajpai1975,ackland2017,hutcheon2019,Liu1999,yao2009}. The dashed lines are linear fits; each have a slope of $\sim-$120 K/meV.}
\label{Fig2a}
\end{figure}

So what is responsible for the wide ranges of theoretical transition temperatures?  A large part of the discrepancies comes from variations in calculated values of $A(0)$, the Helmholtz free energy at zero temperature.  The sensitivity of $T_{\rm bcc}^{\rm \, cp}$ to $\Delta A(0) = A_{\rm cp}(0) - A_{\rm bcc}(0)$ is illustrated in Fig.~\ref{Fig2a}, where we plot theoretical values \cite{schneider1970,bajpai1975,ackland2017,hutcheon2019,Liu1999,yao2009} of $T_{\rm bcc}^{\rm \, fcc}$ and $T_{\rm bcc}^{\, 9R}$ vs $\Delta A(0)$ for Li.  As can be inferred from the linear-fit lines to the two sets of data, a change in $\Delta A(0)$ of 1 meV results in a change in $T_{\rm bcc}^{\rm \, cp}$ of $\sim \! -$120 K.  Therefore computational accuracy in $T_{\rm bcc}^{\rm \, cp}$ of 10 K requires $\Delta A(0)$ to be accurate to $\sim$0.1 meV.

Obtaining such accuracy is a significant challenge, even for first-principles theory (FPT).\footnote{Theories that we consider to be FPT include (i) early pseudopotential calculations that include exchange/correlation and (ii) later density-functional theory that is either all-electron or pseudopotential based.}  Calculation of $\Delta A(0)$ between two phases requires calculation of  $\Delta U_{\rm el}(0) = U_{\rm el}^{\rm  cp}(0) - U_{\rm el}^{\rm  bcc}(0)$, where $U_{\rm el}^{\rm \, cp}(0)$ and $U_{\rm el}^{\rm \, bcc}(0)$ are the electronic contributions to the ground-state energy of cp and bcc phases, respectively.  Energies involved in calculating $U_{\rm el}(0)$ are on the order of $10^2$ to $10^3$ eV.  Therefore, computational accuracy on the order of 1 part in $10^6$ to $10^7$ is required for $\sim$10 K accuracy in $T_{\rm bcc}^{\rm \, cp}$.  Historically, such accuracy has been elusive. For example, theoretical values of $\Delta U_{\rm el}(0)$ between Li fcc and bcc phases range from $-12.2$ meV \cite{BoettgerPRB1985} to $-0.1$ meV \cite{krashaninin1993}.  The variations in FPT energy differences can arise arise from a variety of factors.  For example, for all-electron (AE) calculations the choice of basis set is significant, with Gaussian-type basis orbitals typically yielding larger energy differences than cellular basis sets \cite{trickey2016}.  Energy differences are also sensitivity to the choice of exchange-correlation functional \cite{almeida2003,hanfland2000,jerabek2022}.

Zero-point energy $U_{\rm vib}(0)$ associated with the vibrational ground state of the lattice also contributes to $A(0)$.  As with $\Delta U_{\rm el}(0)$, significant spread in calculated values of $\Delta U_{\rm vib}(0)$ between cp and bcc phases also exist.  For Li fcc and bcc phases, FPT calculated values of $\Delta U_{\rm vib}(0)$ range from 0.4 meV \cite{Liu1999} to 1.7 meV \cite{bajpai1975}.

Here we present an alternative approach to assessing $\Delta A(0) = \Delta U_{\rm el}(0) + \Delta U_{\rm vib}(0)$.  In short, we employ theoretical vibrational spectra from the alkali metals in bcc and cp structures in conjunction with $T_{\rm bcc}^{\rm cp}$ values inferred from experimental data to determine $\Delta A(0)$ and its components $\Delta U_{\rm el}(0)$ and $\Delta U_{\rm vib}(0)$.  In detail, we begin with an embedded-atom-method (EAM) model that is specifically designed to accurately describe vibrations in bcc Li, Na, K, and Rb (Sec.~\ref{SecII}).  The EAM model is then used to calculate vibrational structure of fcc, hcp, and 9$R$ phases of each of these metals (Sec.~\ref{SecIII}).  Utilizing vibrational densities of states (DOS) $g(\omega)$ obtained from our calculations, we directly find values of $\Delta U_{\rm vib}(0)$ between the cp and bcc phases.  The DOS $g(\omega)$ also allows us to calculate the temperature dependence of the Helmholtz energy $A(T)$ for each phase.  By setting $A_{\rm cp}(T_{\rm bcc}^{\rm cp}) = A_{\rm bcc}(T_{\rm bcc}^{\rm cp}) $ at values of $T_{\rm bcc}^{\rm cp}$ inferred from experiment we thence determine $\Delta A(0)$ and $\Delta U_{\rm el}(0)$ (Sec.~\ref{SecIV}).  For K and Rb (which do not exhibit any temperature induced transitions) we are able to infer constraints on $\Delta U_{\rm el}(0)$.

\section{The EAM Model}
\label{SecII}

In the EAM formalism the potential energy $\Phi$ of the solid is a sum of atomic pair potentials $\phi(r_{ij})$ and embedding energies $F(\rho_i)$,
\be{\eq}
\label{1}
\Phi = \frac{1}{2} \sum_{ij} \phi(r_{ij}) + \sum_{i} F(\rho_i).
\en{\eq}
Here $i$ and $j$ ($i \ne j$) label the atoms, $r_{ij}$ is the distance between atoms $i$ and $j$, and $\rho_i$ is the electron (number) density at the position of atom $i$ associated with all other atoms in the solid.  The density $\rho_i$ is assumed to be a sum of atomic densities $f(r_{ij})$,
\be{\eq}
\label{2}
\rho_i = \sum_{j} f(r_{ij}).
\en{\eq}
Specifying the three functions $\phi(r)$, $F(\rho)$, and $f(r)$ defines any particular EAM model.

The EAM model we introduce here is a slight extension of the model developed by Wilson and Riffe (WR) that accurately describes vibrations of alkali metals in the bcc structure \cite{Wilson2012}.   The three defining functions are given by 
\be{\eq}
\label{3}
\phi(r) = \sum_{n=0}^7 K_n \big( r/r_1 -1 \big)^{\! n} \exp \! \Big( \! -n \alpha \big( r/r_1 -1 \big)^{\! 2} \Big),
\en{\eq}
\be{\eq}
\label{4}
F(\rho) = - \big(E_{ \rm coh} - E_{1 \rm v} \big) \big[ 1 - \lambda \, {\rm ln}  ( \rho/\rho_e ) \big] ( \rho/\rho_e )^{\lambda},
\en{\eq}
and
\be{\eq}
\label{5}
f(r) = f_1 \, {\rm exp} \big( \!- 6 ( r/r_1 -1 ) \big).
\en{\eq}
(The pair potential and embedding-energy functions were first introduced by Wang and Boercker \cite{WBJAP1995} and Johnson \cite{JOJMR1989,JohnsonPRB1988}, respectively.)  The only difference between the form of the current model and that used by WR is the introduction of an extra term ($n = 7$) in the pair-potential $\phi(r)$.  In Eqs.~(\ref{3})--(\ref{5}), $r_1 = \sqrt{3/4} \, a_0$ is the bcc-structure nearest-neighbor distance ($a_0$ is the lattice constant), $E_{\rm coh}$ is the cohesive energy, and $E_{\rm 1v}$ is the unrelaxed vacancy formation energy.  We use values for these three quantities from experimental measurements (see Table~\ref{table1}).  We note the value of $f_1$ in Eq.~(\ref{5}) is irrelevant, owing to the embedding energy being a function of $\rho/\rho_e$, where $\rho_e$ is the density when the lattice is at equilibrium.  The pair potential expression given by Eq.~(\ref{3}) is assumed to vanish at distances $r > 1.75 \, a_0$.  This choice results in $\phi(r)$ being nonzero out to the 5th shell for the bcc, hcp, and 9$R$ structures, and out to the 3rd shell for the fcc structure.

In the WR study \cite{Wilson2012} the pair-potential parameters $K_n$ were uniquely determined via seven algebraic equations that involve $E_{\rm 1v}$, the elastic constants $G = (C_{11} - C_{12} + 3 \, C_{44}) / 5$ and $C'$ (experimental values also displayed in Table~\ref{table1}), and three zone-edge phonon frequencies \cite{WBJAP1995}.  Values for the two embedding-energy parameters $\alpha$ and $\lambda$ were selected by visually comparing calculated and experimental dispersion curves.  Surface relaxation was also considered in choosing a value of  $\lambda$ for each metal.

\begin{table}[t]
\caption{Values of experimental input parameters used to construct the alkali-metal EAM models.\label{table1}}
\vspace{0.2cm}

\begin{tabular}{l@{\hspace{0.25cm}}d@{\hspace{0.25cm}}d@{\hspace{0.25cm}}d@{\hspace{0.25cm}}d@{\hspace{0.3cm}}r@{\hspace{0.4cm}}r@{\hspace{0.4cm}}r@{\hspace{0.3cm}}r@{\hspace{0.3cm}}r@{\hspace{0.3cm}}r@{\hspace{0.3cm}}r@{\hspace{0.3cm}}r@{\hspace{0.3cm}}r}

\hline
\hline

\rule{0pt}{2.5ex}& \rm Li& \rm Na& \rm K& \rm Rb \\

\hline

\rule{0pt}{2.5ex}$M$ (amu)\footnote{Mass values correspond to samples measured by neutron scattering: Li [\onlinecite{SmithNIS1968}], Na [\onlinecite{WoodsPR1962}], K [\onlinecite{CowleyPR1966}], and Rb [\onlinecite{CopleyCJP1973}]; Li is isotopically pure; the other metals exhibit natural isotopic abundance.}	&7.00	&23.0	&39.1	&85.5\\

$E_{\rm coh}$ (eV)\footnote{Cohesive energies are from [\onlinecite{Kittel2005}]. } 	&1.63	&1.113	&0.934	&0.852\\

$E_{\rm 1v}$ (eV)\footnote{Vacancy formation energies are from \cite{MacDonaldJCP1953} (Li),  \cite{adlhart1975} (Na), \cite{MundyPRB1971} (K), \cite{MartinPR1965} (Rb). }	&0.40	&0.36	&0.35	&0.30\\

$a_0$ (nm)\footnote{Lattice constants are from \cite{AndersonPRB1985} (Li), \cite{SiegelPR1938,BarrettAC1956} (Na), \cite{SchoutenPRB1974} (K), \cite{BarrettAC1956} (Rb). }	&0.348	&0.424	&0.524	&0.559\\

$G$ (100 Mbar)\footnote{Elastic constants are from \cite{SlotwinskiJPCS1969} (Li),  \cite{QuimbyPR1938,MartinsonPR1969}  (Na), \cite{MarquardtJPCS1965} (K), \cite{GutmanJPCS1967} (Rb). \label{elastic} }		&6.85	&3.79	&1.86	&1.42\\

$C'$ (100 Mbar)\textsuperscript{\ref{elastic}}	&1.13	&0.72	&0.37	&0.27\\

\hline

\end{tabular}
\end{table}

\begin{table*}[t]
\caption{Fitted values of the EAM-model parameters $K_n$, $\alpha$, and $\lambda$.\label{table2}}
\vspace{0.2cm}

\begin{tabular}{l@{\hspace{-0.2cm}}d@{\hspace{-0.2cm}}d@{\hspace{-0.2cm}}d@{\hspace{-0.2cm}}d@{\hspace{0.3cm}}r@{\hspace{0.4cm}}r@{\hspace{0.4cm}}r@{\hspace{0.3cm}}r@{\hspace{0.3cm}}r@{\hspace{0.3cm}}r@{\hspace{0.3cm}}r@{\hspace{0.3cm}}r@{\hspace{0.3cm}}r}

\hline
\hline

& \rm Li& \rm Na& \rm K& \rm Rb \\

\hline

\rule{0pt}{2.5ex}$K_0$	&-0.0604429775	&-0.0531748765	&-0.0510733606	&-0.0491011717   \\

$K_1$ 	&-0.1221933704	&-0.0812858737	&-0.1028668895	&-0.1179195388   \\

$K_2$	&2.0046777475	&1.5904450830	&1.6053249184	&1.6009616092   \\

$K_3$	&-6.5477195147	&-3.9837761136	&-4.6150715295	&-4.2471006924   \\

$K_4$	&12.4561144802	&3.1373244334	&6.9063432497	&5.1462848297   \\

$K_5$	&-16.0838649315	&0.9927369331	&-6.9093478530	&-3.1482335793   \\

$K_6$	&12.3511739623	&-2.5893584833	&4.5956406720	&0.9028679099   \\

$K_7$	&-4.0181864343	&0.9831489102	&-1.4398059614	&-0.0872715445   \\

$\alpha$	&0.17			&0.22			&0.20	&0.04   \\

$\lambda$		&0.256504	&0.425030		&0.531433	&0.545806   \\

\hline
\hline

\end{tabular}
\end{table*}

\begin{figure*}[t]
\centerline{\includegraphics[scale=1]{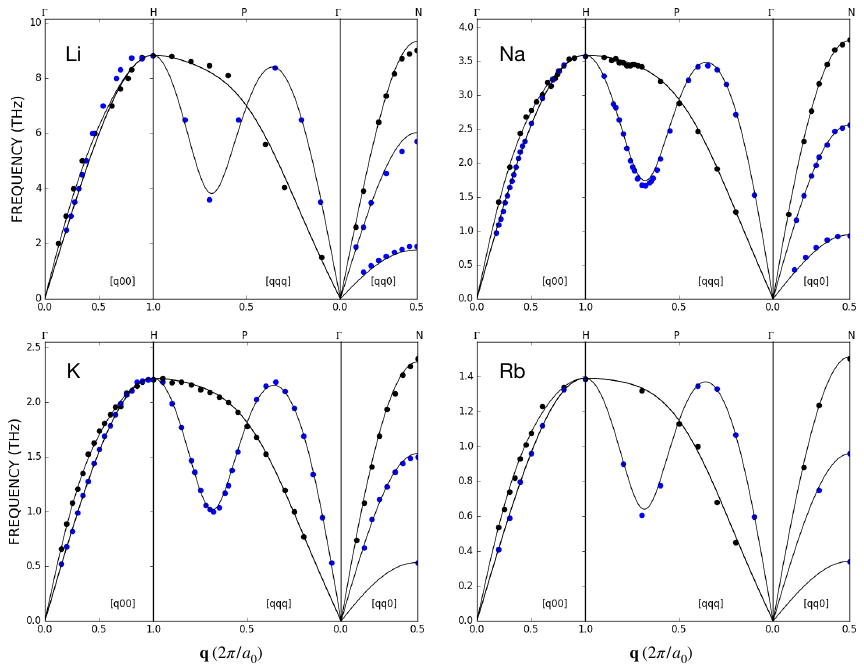}}
\caption{Phonon dispersion curves of Li, Na, K, and Rb in the bcc structure.  Circles are from neutron-scattering experiments: Li [\onlinecite{SmithNIS1968}], Na [\onlinecite{WoodsPR1962}], K [\onlinecite{CowleyPR1966}], and Rb [\onlinecite{CopleyCJP1973}]. Solid curves are best-fit curves to our EAM model. }
\label{Fig1}
\end{figure*}

Our approach in setting values of the model parameters for the four alkali metals studied here builds upon the approach of WR.  First, we constrain our EAM model using the four equations---of the above mentioned seven---that do not involve the zone-edge frequencies,
\begin{equation}
\label{6}
0 = 4 r_1 \phi_1' + 3 r_2 \phi_2' + 6 r_3 \phi_3' +12 r_4 \phi_4' +4 r_5 \phi_5',
\end{equation}
\begin{equation}
\label{7}
15 \Omega G = 4 r_1^2 \left( \phi_1'' + \phi_2'' + 4 \phi_3'' + 11 \phi_4'' + 4 \phi_5''   \right),
\end{equation}
\begin{equation}
\label{8}
3 \Omega C' = 2 r_1^2 \left( \phi_2'' - \frac{\phi_2'}{r_2} +  \phi_3'' - \frac{\phi_3'}{r_3} + \frac{64}{11} \phi_4'' - \frac{64}{11} \frac{\phi_4'}{r_4}   \right),
\end{equation}
and
\begin{equation}
\label{9}
E_{1 \rm v} = - 4 \phi_1 - 3 \phi_2 - 6 \phi_3 - 12 \phi_4 - 4 \phi_5.
\end{equation}
Here $\Omega = a_0^3/2$ is the bcc equilibrium atomic volume, and  $r_i$ is the distance to the $i$th shell of neighboring atoms.  The quantities $\phi'_i$ and $\phi''_i$ are derivatives of the pair potential evaluated at the equilibrium value of shell distance $r_i$.  We also employ one more equation of constraint,
\begin{equation}
\label{10}
0 = 2 r_{1 \rm f} \phi'_{1 \rm f} + r_{2 \rm f} \phi'_{2 \rm f} + 4 r_{3 \rm f} \phi'_{3 \rm f}, 
\end{equation}
with the condition $r_{1 \rm f} = 2^{-1/6} \, a_0$.  Here $r_{i \rm f}$ is the $i$th-shell distance in the fcc structure.  This constraint ensures the fcc and bcc equilibrium atomic volumes are (essentially) the same.\footnote{If the pair-potential term in Eq.~(\ref{1}) were the only contribution to the energy $\Phi$, then this constraint would ensure exact equality of the fcc and bcc equilibrium atomic volumes.  Owing to slight variations in the embedding energy $F(\rho)$ among the different structures, the equilibrium volumes are not identical, but they differ by an insignificant amount ($\Delta \Omega / \Omega < 0.75$\% for all four metals). }  This equal-volume constraint is motivated by both experimental measurements \cite{barrett1947,owen1954,BarrettAC1956,smith1987, berliner1989,berliner1992,smith1991} and FPT energy calculations \cite{almeida2003,wang2004,kulkarni2011,ackland2017,hutcheon2019,jerabek2022,gaissmaier2020} of alkali metals in bcc and close-packed structures, which indicate negligible differences ($\lesssim 1$\%) in equilibrium values of $\Omega$ among the various structures.  Differing from the WR approach of exactly matching three zone-edge frequencies in the determination of the $K_n$'s, here we least-squares fit the EAM model [with the constraints of Eqs.~(\ref{6})--(\ref{10}) imposed] to all available experimental vibrational frequencies (in a given study) along the three high-symmetry directions [(100), (110), and (111)] in the bcc structure.  Those experimental data \cite{SmithNIS1968,WoodsPR1962,CowleyPR1966,CopleyCJP1973} are displayed in Fig.~\ref{Fig1}. In the least-squares analysis each phonon branch is given equal weighting, irrespective of the number of experimental frequencies reported in that branch.  Along with the $K_n$'s, values for the two remaining parameters $\alpha$ and $\lambda$ are also determined.  Fitted values of these parameters for each metal are displayed in Table \ref{table2}.  The least-squares-fit dispersion curves are also displayed in Fig.~\ref{Fig1}.  As is evident there, our EAM calculated curves match the experimental data quite well. 

Compared to the prior WR EAM calculations \cite{Wilson2012}, the present model does a slightly better job at describing the bulk vibrational structure of the alkali metals.  The major advantage of the present approach, however, is the implementation of least-squares fitting, which effects a more objective determination of the model's parameters.

In passing, we note our dispersion curves are significantly more accurate than those previously calculated for the alkali metals with most other EAM models \cite{GAJMR1992,CMPRB1998,HuMSMSE2002,XieCJP2008,ZhangJLTP2008,nichol2016,ko2017,kim2020,qin2022} and even FPT \cite{ko2017,ackland2017,hutcheon2019,kim2020}.  On the whole, EAM calculations exhibit no general trend in deviations from experimental frequencies.  Conversely, for Li at least, FPT calculations tend to overestimate phonon frequencies by $\sim \,$6\%. \cite{ko2017,ackland2017,hutcheon2019}.

\section{Vibrations in Close-Packed Structures}
\label{SecIII}

Using our EAM model we now investigate vibrations of Li, Na, K, and Rb arranged in fcc, hcp, and 9$R$ lattices.  Vibrational frequencies for these three structures (as well as bcc) are found using general equations for an EAM dynamical matrix \cite{riffe2018}.  In these calculations the atomic volume in each cp structure is constrained to equal the bcc equilibrium atomic volume $\Omega$.  We also assume ideal stacking ($c = \sqrt{8/3} \, a$) of the cp planes in the hcp and 9$R$ structures.  In turn, we discuss dispersion relations along high-symmetry directions, vibrational densities of states (DOS), and moment Debye temperatures $\Theta_m$, with particular emphasis on moment $m = 1$. 

\begin{figure*}[t]
\centerline{\includegraphics[scale=0.73]{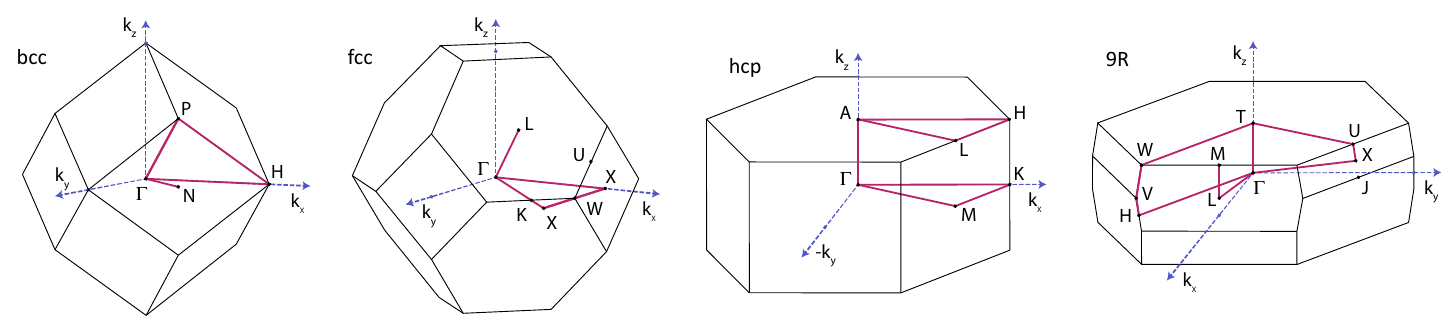}}
\caption{First Brillouin zones for bcc, fcc, hcp and $9R$ structures.  The line segments between high-symmetry points indicate the $k$-space paths along which dispersion curves in Fig.~\ref{Fig3} are plotted.}
\label{Fig2}
\end{figure*}

Relevant to the these calculations---and especially the dispersion-curve plots to follow---in Fig.~\ref{Fig2} we illustrate first Brillouin zones (BZ's) for the four structures.  The labels we use for the high-symmetry points in the bcc, fcc, and hcp structures are standard notation.  For the 9$R$ structure there does not appear to be a canonical set of labels.  We adopt the detailed scheme of Yorikawa and Muramatsu \cite{yorikawa2008}.  We note the points M and U are equivalent.

\begin{figure*}[t]
\centerline{\includegraphics[scale=1.50]{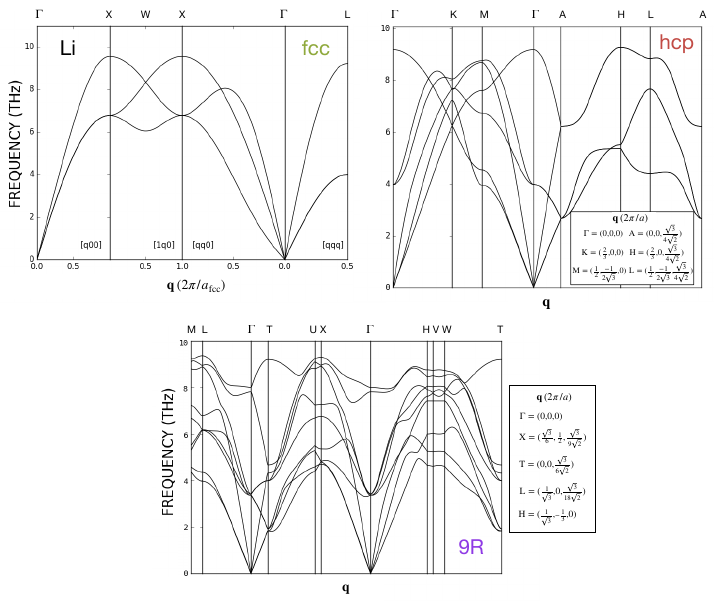}}
\caption{Phonon dispersion curves of Li in close-packed fcc, hcp, and $9R$ structures.  Here $a_{\rm fcc}$ is the fcc lattice constant, and $a$ is the hcp/9$R$ nearest-neighbor distance.  Locations of the high-symmetry BZ points along the top of each panel are indicated in Fig.~\ref{Fig2}. }
\label{Fig3}
\end{figure*}

Dispersion curves for Li along high-symmetry directions in the close-packed structures are illustrated in Fig.~\ref{Fig3}.  Owing to the similarity of analogous curves for Na, K, and Rb, the cp dispersion curves for these three metals are not presented here.  Such similarity is not surprising, given the congruency of bcc dispersion curves for these metals.  The increasing complexity of the dispersion-curve plots with the progression from fcc to hcp to 9$R$ is the result of the number of atoms per unit cell progressing from one to two to three in these structures.

\begin{figure}[t]
\centerline{\includegraphics[scale=0.6]{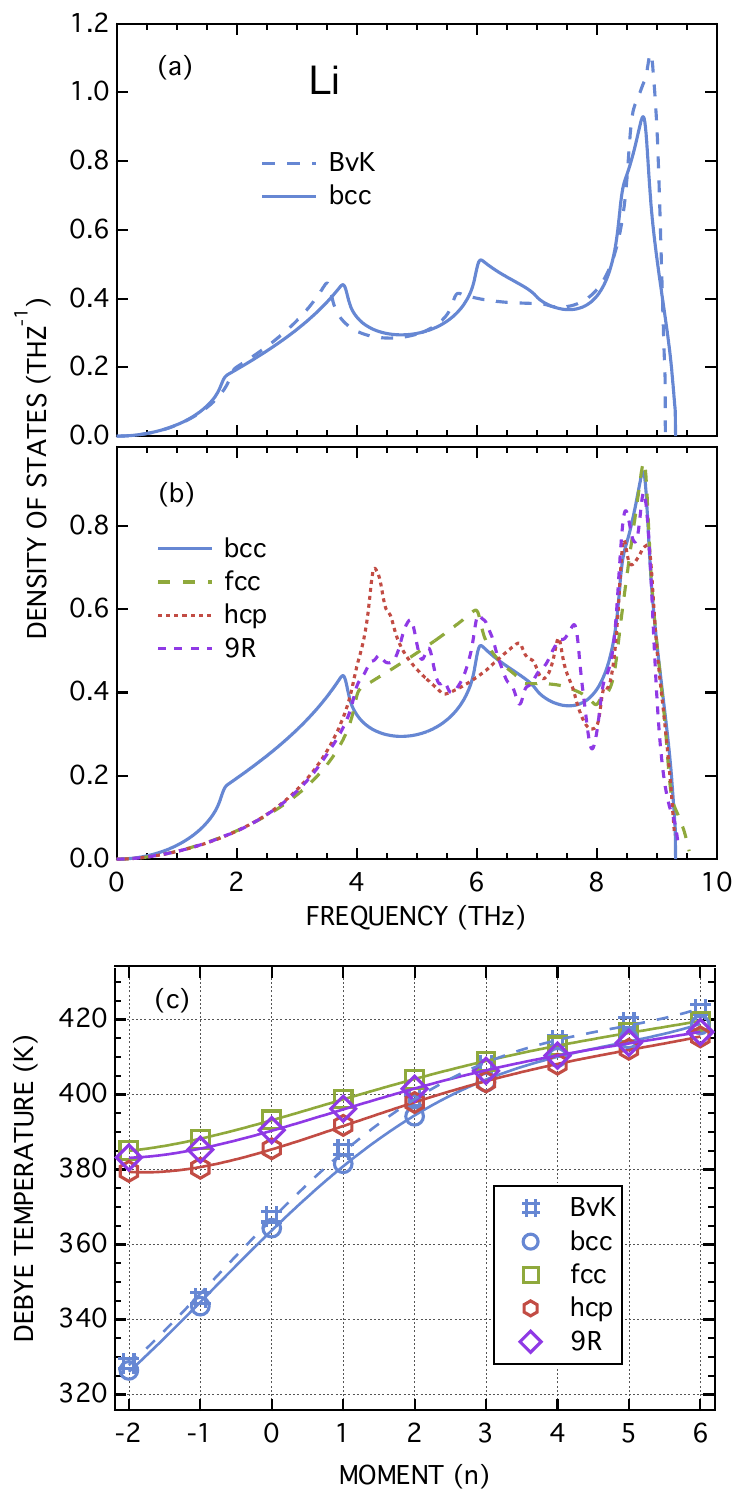}}
\caption{Li vibrational densities of states (DOS) and moment Debye temperatures $\Theta_m$ in bcc and close-packed fcc, hcp, and $9R$ structures.  DOS for the bcc structure are obtained using both experimental BvK force constants and our EAM model.}
\label{Fig5}
\end{figure}

Vibrational DOS for Li in the four structures are displayed in panels (a) and (b) of Fig.~\ref{Fig5}.  These DOS were calculated by finding vibrational modes and their corresponding frequencies for at least $10^6$ points within the irreducible section of the first BZ for each structure.  For the bcc structure we also calculate the DOS using Born-von-K\'arm\'an (BvK) force constants that have been obtained from previous analysis of the experimental dispersion-curve data shown in Fig.~\ref{Fig1} \cite{SmithNIS1968,WoodsPR1962,CowleyPR1966,CopleyCJP1973}.

As Fig.~\ref{Fig5}(b) shows, the major difference between the bcc and cp-structures DOS is the much larger weight at lower frequencies in the bcc structure.  This low-frequency enhancement is due to modes associated with the elastic constant $C'$, which manifest along [qq0] as the lowest-frequency dispersion curve (which comprises transverse modes polarized in the $\langle 1 1 0 \rangle$ direction).  As first suggested by Barrett in the initial structural study of Li \cite{barrett1947}, entropy associated with these modes primarily drives the martensitic phase transitions in Li and Na.

In Fig.~\ref{Fig5}(c) we plot moment Debye temperatures $\Theta_m$ \cite{Grimvall1981,Riffe2023} for Li in all four structures.  For reference, these Debye temperatures are calculated from the vibrational DOS function $g(\omega)$ via
\be{\eq}
\label{B2}
\Theta_m =  \! \frac{\hbar}{k_B} \!  \, \bigg[ \frac{m+3}{3} \frac{\int d\omega \, g(\omega) \, \omega^m}{\int d\omega \, g(\omega) }\bigg]^{1/m}
\en{\eq}
($m = -2, -1, 1, 2, 3, ...$) and 
\be{\eq}
\label{B3}
\Theta_0 = \frac{\hbar \omega_0}{k_B} \, \exp \! \bigg( \frac{1}{3} +  \frac{\int d\omega \, \ln(\omega/\omega_0) g(\omega)}{\int d\omega \, g(\omega) } \bigg)
\en{\eq} 
($m = 0$).  Here $\omega_0$ is an arbitrary frequency.  The much lower values of $\Theta_m$ for $m = -2$, $-1$, and 0 in the bcc structure is related to the aforementioned larger DOS at low frequencies (small $\omega$) for this structure.  Conversely, the close correspondences of $\Theta_m$ for $m \ge 3$ derives from the similar overall structure of $g(\omega)$ at larger values of $\omega$ for all four phases.  We note that although the BvK and EAM model calculated Li DOS have significant deviations, the moment Debye temperatures differ by only $\sim1$\% for all values of $m$.  Similarly good---if not better---agreement is exhibited by this measure for Na, K, and Rb.

\begin{figure}[t]
\centerline{\includegraphics[scale=0.50]{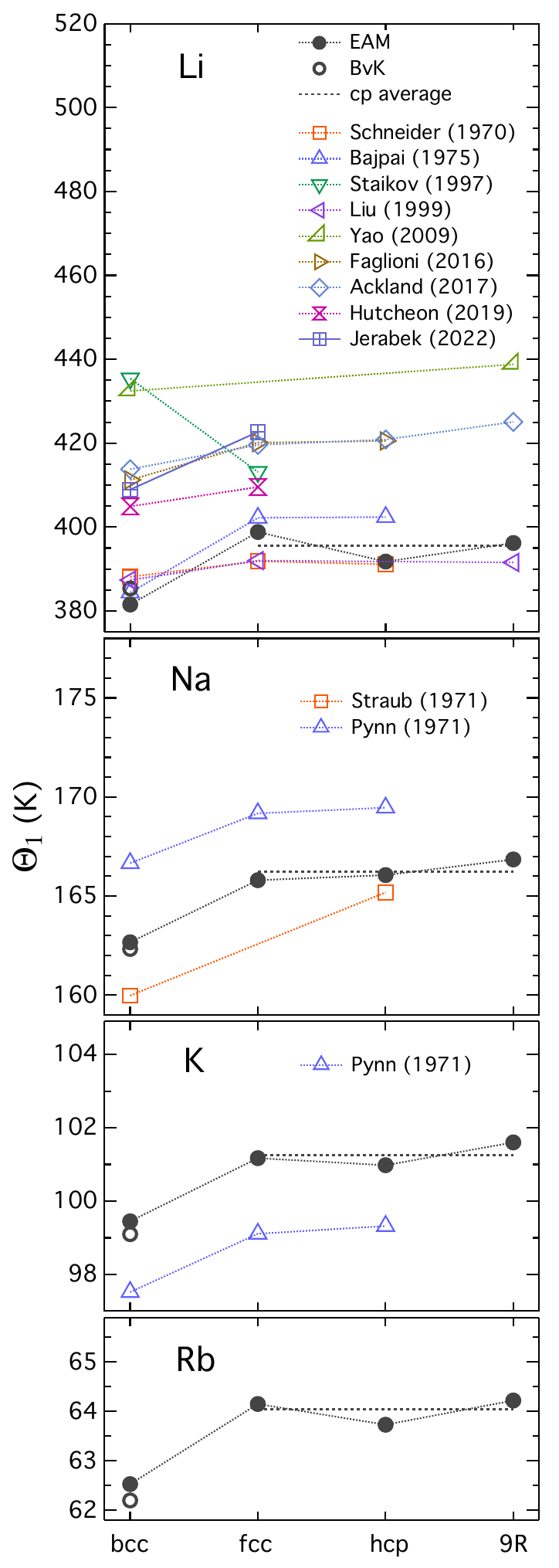}}
\caption{Values of moment Debye temperature $\Theta_1$ for Li, Na, K, and Rb obtained from theoretical calculations in bcc and cp structures.  EAM refers to present results, BvK refers to $\Theta_1$ calculated from BvK force constants (extracted from experimental data), and cp average is an average of cp-structure $\Theta_1$ values from the present EAM calculations.  Other results have been obtained from the literature \cite{schneider1970,bajpai1975,Liu1999,yao2009,faglioni2016,ackland2017,hutcheon2019,jerabek2022,straub1971,pynn1971}.}
\label{Fig6}
\end{figure}

Because $\Theta_1$ is proportional to the zero-point vibrational energy
\be{\eq}
\label{B4}
U_{\rm vib}(0) = \frac{9}{8} \, k_B \Theta_1,
\en{\eq} 
 this particular Debye temperature is especially key with regard to the phase transitions of Li and Na.  We are thus motivated to compare $\Theta_1$ among the different structures for all four metals; this comparison can be found in Fig.~\ref{Fig6}.  We observe that $\Theta_1$ for the cp structures is consistently larger than that for the bcc phase for all four metals.  On average, our cp $\Theta_1$ values are larger than bcc values by 2.5 $\pm$ 0.3\%.  In terms of zero-point energies, the increase in $U_{\rm vib}(0)$ (from bcc to cp phases) is characterized by energy differences of $1.35 \pm 0.34$, $0.35 \pm 0.05$, $0.17 \pm 0.03$, and $0.15 \pm 0.03$ meV for Li, Na, K, and Rb, respectively.

We have perused the literature for previous investigations that compare bcc and cp alkali-metal vibrations.  We located nine previous studies on Li \cite{schneider1970,bajpai1975,staikov1997,Liu1999,yao2009,faglioni2016,ackland2017,hutcheon2019,jerabek2022}, two on Na \cite{straub1971,pynn1971}, one on K \cite{pynn1971}, but none on Rb.  All of these calculations are pseudopotential (PP) based.  For comparison with our results, we have extracted values for $\Theta_1$ from these studies. Those values are also displayed in Fig.~\ref{Fig6}.    Aside from the outlying result of Staikov \textit{et al.} \cite{staikov1997}, all calculations---in agreement with ours---exhibit larger values of $\Theta_1$ for the cp structures.  We note the following details for Li.  The values of bcc $\Theta_1$ from the more recent calculations (2016 and later) are systematically larger (by $\sim$6\%) than those from the earlier calculations, with the earlier values of bcc $\Theta_1$ being in much better agreement with our calculations and with $\Theta_1$ derived from BvK force-constant (FC) analysis of experimental data.  The difference in $\Theta_1$ between cp and bcc structures varies rather dramatically among the PP calculations, with the differences ranging from 3.7 K \cite{schneider1970} to 17.8 K \cite{bajpai1975}.  Expressed in terms of $\Delta U_{\rm vib}(0)$, this difference range is 0.36 meV to 1.72 meV.  Clearly, no consensus exists among the PP calculations with regard to the increase in zero-point energy for Li.  Previous calculations that are most in line with our results are those of Bajpai \textit{et al}. \cite{bajpai1975} for Li and Pynn \textit{et al}. \cite{pynn1971} for Na and K.

\section{BCC to CP Phase Transitions}
\label{SecIV}

As discussed above, previous efforts to understand details of the martensitic phase transitions in Li and Na focus on differences in the free energy $A(T)$ among the bcc and cp structures.  In principle, $A(T)$ has temperature-dependent has contributions from both electronic and vibrational excitations,
\be{\eq}
\label{001}
A = U_{\rm el} + U_{\rm vib} - T \, (S_{\rm el} +S_{\rm vib}).
\en{\eq}
However, due to the relatively low temperatures involved, several approximations to $A(T)$ can be profitably made without the introduction of significant error.  First, the contributions from electronic excitations can be neglected, leaving only the ground-state electronic energy $U_{\rm el}(0)$ to contribute to the electronic component of $A(T)$ \cite{Liu1999,jerabek2022}.  Second, because vibrational anharmonicity is minimal at the relatively low temperatures of interest, we can treat the phonon contribution in the harmonic approximation \cite{yao2009,hutcheon2019}.  Utilizing these two approximations, the free energy can be expressed as \cite{pynn1971,straub1971,ackland2017,McQuarrie1976}
\begin{equation}
\label{002}
A(T) = U_{\rm el}(0)  +  A_{\rm vib}(T)
\end{equation}
where
\begin{align}
\label{003}
A_{\rm vib}(T) = & \,\, U_{\rm vib}(0)  \nonumber \\ 
				&+ \! \int \!\! d\omega \, g(\omega) \, k_B T \ln \!\big( 1 - e^{-\hbar \omega / k_B T}   \big).
\end{align}

\noindent We note the integral term in Eq.~(\ref{003}) is the combination $\delta U_{\rm vib}(T) - T \, S_{\rm vib}(T)$, where $\delta U_{\rm vib}(T) = U_{\rm vib}(T) - U_{\rm vib}(0)$.

As described above in reference to Fig.~\ref{Fig1a}, cp phases appear in Li and Na as the temperature descends from RT.  Therefore, $\Delta A(T)$ for the exhibited cp phases becomes negative at sufficiently low temperature.  We thus ask, what is the transition temperature $T^{\rm cp}_{\rm bcc}$ for each of these phases?

Let's first consider Li.  Several independent experiments imply $T^{\rm fcc}_{\rm bcc}$ lies within the range 155 K to 180 K (illustrated by the red band in the Li panel in Fig.~\ref{Fig1a}), where the martensite transforms back to bcc upon heating \cite{smith1987,schwarz1990,schwarz1991,smith1994,maier1995,krystian2000}.  First, just below the transition back to bcc only the fcc phase of the martensite is still present \cite{schwarz1990,schwarz1991,pichl2003}.  Second, as we see below when considering $\Delta A(T)$, a value of $T^{\rm fcc}_{\rm bcc}$ within the range 155 K to 180 K implies that fcc is the true ground-state phase of Li.  That is, it has the lowest value of $A(0)$.  Such ground-state identification is consistent with the observation that cold working transforms a sample initially cooled to 77 K to the fcc phase \cite{barrett1947}.  That fcc is the ground state is also supported by recent x-ray diffraction measurements of Li, where fcc is the only cp phase observed if the sample is first pressurized (to 10 GPa), the temperature is subsequently lowered to 10 K, and then the pressure is largely released (to 0.5 GPa for Li$^6$ and 2.5 GPa for Li$^7$) \cite{ackland2017}.  Finally, thermal-cycling measurements directly show that the fcc phase, once formed, is stable down to at least 10 K \cite{pichl2003}.  We thus conclude $T^{\rm fcc}_{\rm bcc} = 167.5 \pm 12.5$ K.

This determination leaves us to deduce $T_{\rm bcc}^{9 \! R}$, the transition temperature between the bcc and 9$R$ phases.  It has already been suggested that $T_{\rm bcc}^{9 \! R}$ is congruent with the temperature that the martensite first forms upon cooling \cite{ackland2017}, which is in the range of 65 K to 85 K \cite{ackland2017} (as indicated by the blue band in the Li panel of Fig.~\ref{Fig1a}).  We thus assume $T_{\rm bcc}^{9 \! R} = 75 \pm 10$ K.  This assignment of $T_{\rm bcc}^{9 \! R}$ is also consistent with a thermal cycling experiment, where a sample with an already formed and stable fcc fraction is repeatedly cycled between 10 K and 120 K. This temperature cycling results in reproducible transition back and forth between bcc and 9$R$ in (minority) non-fcc regions of the sample, with the transition back to bcc occurring at $\sim$90 K, just above the 65 K to 85 K region \cite{pichl2003}.

\begin{figure*}[t]
\centerline{\includegraphics[scale=1.33]{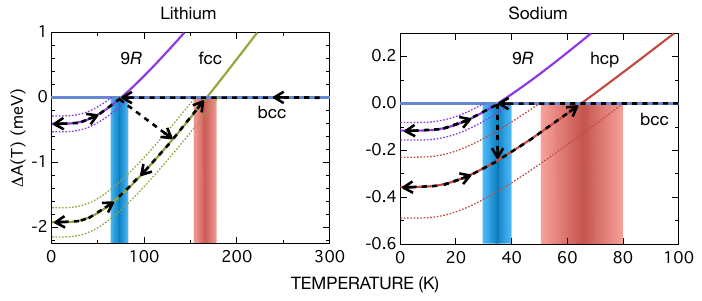}}
\caption{Helmholtz free-energy differences $\Delta A(T)$ (with respect to the bcc phase) vs temperature $T$ for Li and Na.  Dashed arrows schematically illustrate the observed sequence of structural phases when starting at high temperatures and finally ending at the transition back to the pure bcc phase. (Although not indicated with the arrows, in both systems some fraction remains in the bcc phase at all temperatures.)  The blue and red bands---which indicate the temperature ranges over which the indicated transitions are experimentally observed---correspond to the same bands in Fig.~\ref{Fig1a}.  The dotted curves illustrate uncertainties in $\Delta A(0)$ that arise from the ranges of experimentally observed transition temperatures.}
\label{Fig8}
\end{figure*}

\begin{table*}[t]
\caption{Zero temperature energy differences $\Delta A(0)$, $\Delta U_{\rm vib}(0)$, and $\Delta U_{\rm el}(0)$ [$\Delta A(0) = \Delta U_{\rm vib}(0) + \Delta U_{\rm el}(0)$] for Li, Na, K, and Rb.  All energies are with respect to the bcc phase.  Reported uncertainties derive from experimental uncertainties (represented by the widths of the blue and red bands in Figs. \ref{Fig1a} and \ref{Fig8}) in transition temperatures (see text for detail).    \label{table3}}
\vspace{0.2cm}

\begin{tabular}{l@{\hspace{1cm}}l@{\hspace{1cm}}d@{\hspace{1cm}}d@{\hspace{1cm}}d@{\hspace{0.3cm}}r@{\hspace{0.4cm}}r@{\hspace{0.4cm}}r@{\hspace{0.3cm}}r@{\hspace{0.3cm}}r@{\hspace{0.3cm}}r@{\hspace{0.3cm}}r@{\hspace{0.3cm}}r@{\hspace{0.3cm}}r}

\hline
\hline

&  & \rm fcc - bcc& \rm hcp - bcc  & 9R \rm - bcc \\

& $(T = 0)$ & \rm (meV)& \rm (meV)  & \rm (meV) \\

\hline

\rule{0pt}{2.5ex}Li	& $\Delta A$			& -1.92 \pm 0.23	& >0			&  -0.40 \pm 0.13  \\

				& $\Delta U_{\rm vib}$	& 1.68			& 0.99			& 1.40  \\

				& $\Delta U_{\rm el}$ 	& -3.60 \pm 0.23	& > - 0.99			& -1.80 \pm 0.13  \\
				
\rule{0pt}{4ex}Na	& $\Delta A$		& > 0				& -0.36 \pm 0.13	&  -0.12 \pm 0.04  \\

				& $\Delta U_{\rm vib}$	& 0.30			& 0.33			& 0.41  \\

				& $\Delta U_{\rm el}$ 	& >-0.30 			& -0.69 \pm 0.13	& -0.53 \pm 0.04  \\
				
\rule{0pt}{4ex}K		& $\Delta A$			& > 0			& > 0			&  > 0  \\

				& $\Delta U_{\rm vib}$	& 0.17			& 0.15			& 0.21  \\

				& $\Delta U_{\rm el}$ 	& > -0.17			& >-0.15			& > -0.21  \\
				
\rule{0pt}{4ex}Rb		& $\Delta A$			& > 0			& > 0			&  > 0  \\

				& $\Delta U_{\rm vib}$	& 0.16			& 0.12			& 0.16  \\

				& $\Delta U_{\rm el}$ 	& > -0.16			& >-0.12			& > -0.16  \\

\hline
\hline

\end{tabular}
\end{table*}

The combination of (i) our calculations of $A_{\rm vib}(T)$ [via Eq.~(\ref{003})] and (ii) our identifications of $T^{\rm fcc}_{\rm bcc}$ and $T^{9 \! R}_{\rm bcc}$ enables us to infer $\Delta A(T)$ for both cp phases.  In Fig.~\ref{Fig8} we plot $\Delta A(T)$ vs $T$ for the fcc, $9R$, and bcc phases. [Note, trivially  $\Delta A(T) = 0$ for bcc, as all differences are with respect to this phase.]  It is worth keeping in mind that the $\Delta A(T)$ curve with the smallest value (at any given $T$) represents the thermodynamically stable state.  To re-emphasize, at high temperature this is the bcc phase, but below $T^{\rm fcc}_{\rm bcc} = 167.5 \pm 12.5$ K, this is the fcc phase.  Along with the solid curves that represent $\Delta A(T)$ for the fcc and $9R$ phases, we indicate uncertainties in $\Delta A(T)$ via the two sets of dotted curves, which are simply the corresponding $\Delta A(T)$ curves, but (vertically) shifted so that they intersect the lower and upper limits of the transition-temperature regions.  These curves thence provide an estimate of uncertainties in $\Delta A(0)$ that arise from the experimental uncertainties in $T^{\rm fcc}_{\rm bcc}$ and $T^{9 \! R}_{\rm bcc}$.  This analysis yields $\Delta A(T) = -1.92 \pm 0.23$ meV ($-0.40 \pm 0.13$ meV) for fcc ($9R$) Li.

Using the $\Delta A(T)$ curves in Fig.~\ref{Fig8}, we can build a schematic picture of the Li martensitic transitions in $T$--$\Delta A(T)$ space, also illustrated in Fig.~\ref{Fig8}.  With initial cooling from RT, Li bypasses $T_{\rm bcc}^{\rm fcc}$ with no change in structure.  Eventually a 9$R$ martensite forms at $T_{\rm bcc}^{9 \! R} = 75 \pm 10$ K.  There are several theories that describe why bcc Li is able to directly transforms to $9R$, but not fcc \cite{gooding1988,schwarz1991,blaschko1999,pichl2003}.  The cp disorder that simultaneously forms \cite{berliner1986A,smith1987,schwarz1990,schwarz1991,berliner1986B,berliner1989} is apparently a near-surface effect \cite{pichl2003}, and so we shall not consider it further.  Upon further cooling (to at least 4 K \cite{mccarthy1980}), the $9R$ phases persists.  Upon heating, hysteresis becomes apparent:  near 75 K an fcc component emerges from the $9R$ fraction \cite{schwarz1990,schwarz1991,smith1990}.  The fcc phase becomes (almost) entirely dominant by $\sim$130 K.  If the sample is cooled from this point, then the fcc phase remains stable (to at least 10 K) \cite{pichl2003}.  As just mentioned, cycling between 10 K and 120 K induces any remaining non-fcc martensite to transform between 9$R$ and bcc.  Finally, if the sample is warmed to $T^{\rm fcc}_{\rm bcc} = 167.5 \pm 12.5$ K, then it completely reverts back to bcc \cite{smith1987,schwarz1990,schwarz1991,smith1994,maier1995,krystian2000}.

We now turn our attention to transition temperatures $T_{\rm bcc}^{\rm cp}$ for Na.  Similar to Li, the low temperature martensite primarily consists of two phases, in this case 9$R$ and hcp \cite{berliner1989,berliner1992,schwarz1992}.  Also similar to Li, only one of these phases---hcp in the case of Na---is present before final transformation back to bcc (upon heating) \cite{berliner1992,schwarz1992}.  For Na this occurs between 50 K and 80 K (illustrated by the red band in the Na panel in Fig.~\ref{Fig1a}) \cite{BarrettAC1956,szente1988,berliner1992,schwarz1992,abe1994}.  In analogy with $T_{\rm bcc}^{\rm fcc}$ for the parallel fcc-bcc transition in Li, we identify $T_{\rm bcc}^{\rm hcp} = 65 \pm 15$ K.  This identification leaves $T_{\rm bcc}^{9 \! R}$ to be assigned to the lower transition-temperature range of 30 K to 40 K, where the martensite first appears upon cooling (illustrated by the blue band in the Na panel in Fig.~\ref{Fig1a}) \cite{BarrettAC1956,stedman1976,schwarz1992,berliner1992,abe1994,abe1997,blaschko1984,szente1988}.  Hence, for Na  $T_{\rm bcc}^{9 \!R} = 35 \pm 5$ K.

Using these $T_{\rm bcc}^{\rm cp}$ assignments in conjunction with our $\Delta A_{\rm vib}(T)$ calculations we find $\Delta A(0) = -0.36 \pm 0.13$ meV ($-0.12 \pm 0.04$ meV) for hcp ($9R$) Na.  We note the difference between these two energy differences is less than 0.2 meV.

A $T$--$\Delta A(T)$ schematic of the Na transitions is shown in Fig.~\ref{Fig8}.  This picture is similar to that for Li, but there are key differences.  Upon cooling, the higher transition temperature ($T_{\rm bcc}^{\rm hcp}$ for Na) is bypassed with no structural transformation (as is the case with Li), but eventually the martensite is formed at $T_{\rm bcc}^{9 \!R} = 35 \pm 5$ K \cite{BarrettAC1956,stedman1976,schwarz1992,berliner1992,abe1994,abe1997,blaschko1984,szente1988}.  In contrast to Li, both cp phases initially appear.  These phases persist to at least 10 K \cite{szente1988,schwarz1992,abe1994}.  Upon subsequent heating, the return to bcc also differs from Li:  in Li a predominant fraction of the $9R$ phase converts to the more stable fcc structure, but in Na the $9R$ phase simply converts back to bcc.  This conversion happens close to the 30 K to 40 K range \cite{berliner1992,schwarz1992}, which is consistent with our above identification $T_{\rm bcc}^{9 \!R} = 35 \pm 5$ K.   Upon further heating, the hcp fraction reverts to bcc at $T_{\rm bcc}^{\rm hcp} = 65 \pm 15$ K.

\begin{figure}[t]
\centerline{\includegraphics[scale=0.53]{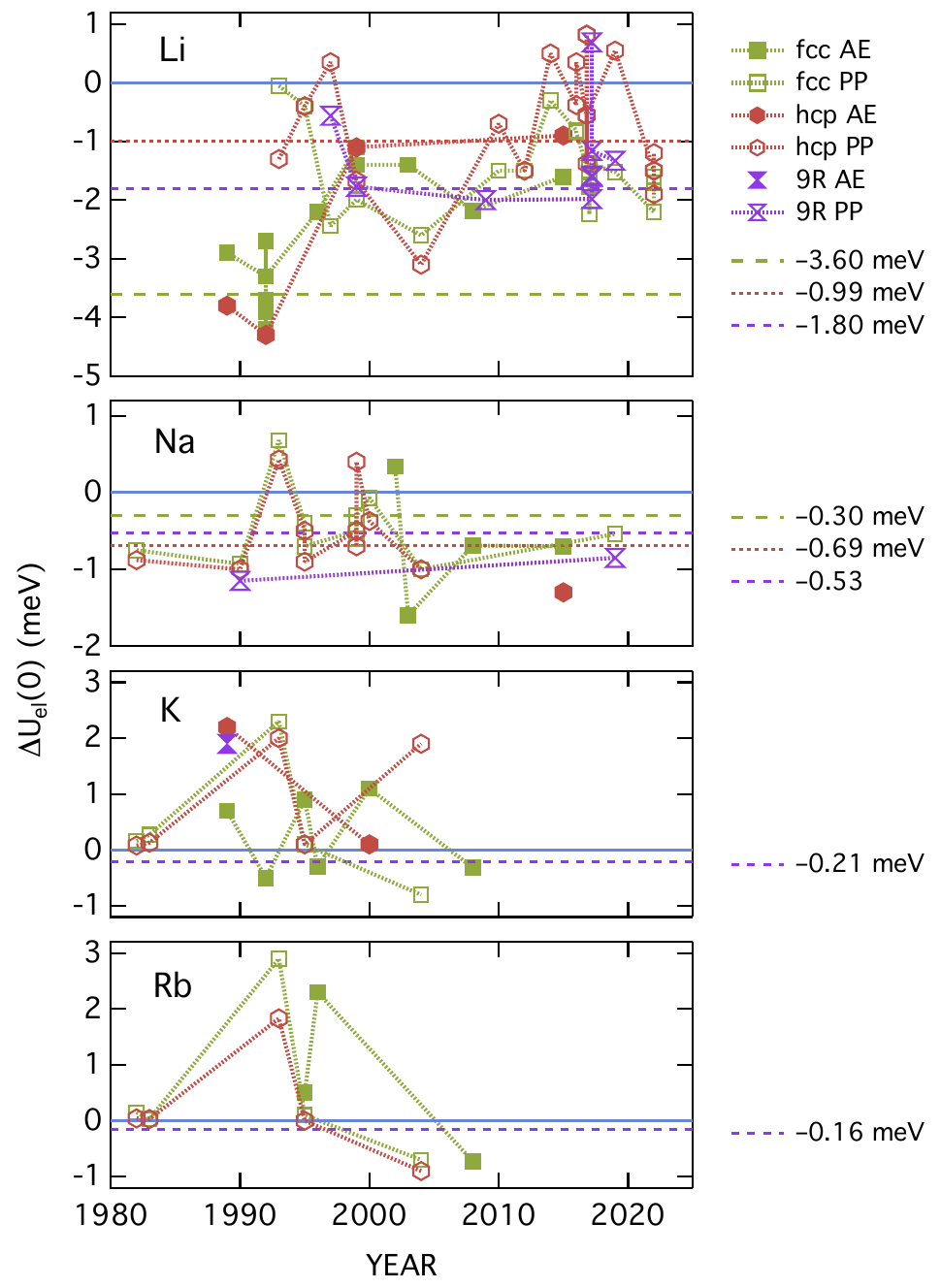}}
\caption{Literature values of electronic energy difference $\Delta U_{\rm el}(0) = U_{\rm cp}(0) - U_{\rm bcc}(0)$ for Li, Na, K, and Rb from first-principles-theory (FPT) calculations published since 1980.  Abbreviations AE and PP refer to all-electron and pseudopotential calculations, respectively.  Horizontal lines correspond to $\Delta U_{\rm el}(0)$ values (or lower limits) given in Table \ref{table3}, with long-dashed green, dotted red, and short-dashed violet lines corresponding to fcc, hcp, and $9R$, respectively.}
\label{Fig9}
\end{figure}

So what is $\Delta U_{\rm el}(0)$ for the relevant phases of Li and Na?  We calculate values for $\Delta U_{\rm el}(0)$ using the $\Delta A(0)$ values just described and $\Delta U_{\rm vib}(0)$ values obtained from our EAM theory:  $\Delta U_{\rm el}(0) = \Delta A(0) - \Delta U_{\rm vib}(0)$.  We find $\Delta U_{\rm el}(T) = -3.60 \pm 0.23$ meV ($-1.82 \pm 0.13$ meV) for fcc ($9R$) Li.  Similarly, $\Delta U_{\rm el}(T) = -0.69 \pm 0.13$ meV ($-0.53 \pm 0.04$ meV) for hcp ($9R$) Na.  In Table \ref{table3} we summarize all $T = 0$ energy differences [$\Delta A(0)$, $\Delta U_{\rm vib}(0)$, and $\Delta U_{\rm el}(0)$] for not only Li and Na, but also for K and Rb (see the discussion below).  

Curiously, the hcp phase is not observed in the Li martensite \cite{berliner1989}.  Similarly, Na does not exhibit the fcc structure \cite{schwarz1992}.  Given that hcp \textit{is} observed in Na and fcc  \textit{is} observed in Li, the simplest explanation for the absent phases is that they are not thermodynamically stable with respect to any of the observed phases.  From this ansatz it follows $\Delta A(0) > 0$ for these two phases, as we have indicated in Table \ref{table3}.  Using our calculated values of $\Delta U_{\rm vib}(0)$, we thence put limits on $\Delta U_{\rm el}(0)$:  $\Delta U_{\rm el}(0) > -0.99$ meV for Li in the hcp configuration and $\Delta U_{\rm el}(0) > -0.30$ meV for fcc Na.  We note that because $\Delta U_{\rm vib}(0)$ is positive, it is certainly possible that $\Delta U_{\rm el}(0)$ is negative for these two absent cp phases, as is the case for the other cp structures.

What about K and Rb?  Insofar as no phases transitions have been observed in these materials, we are also only able to put limits on $\Delta A(0)$ and $\Delta U_{\rm el}(0)$ for these metals.  As with hcp Li and fcc Na, for K and Rb we assume $\Delta A(0) > 0$ for all three cp phases studied here,  which results in the limits on $\Delta U_{\rm el}(0)$ shown in Table \ref{table3}.  Again, it is possible that $\Delta U_{\rm el}(0)$ is negative for the cp phases in K and Rb, but, if so, the the magnitude of $\Delta U_{\rm el}(0)$ must be quite small, on the order of 0.2 meV or less.  

It is worth considering how our results for $\Delta U_{\rm el}(0)$ compare to results from FPT.  To this end, in Fig.~\ref{9} we plot values of $\Delta U_{\rm el}(0)$ from the literature \cite{boettger1989,nobel1992,perdew1992,papa1992,krashaninin1993,fiolhais1995,sliwko1996,staikov1997,Liu1999,doll1999,hanfland1999,almeida2003,wang2004,xie2008A,yao2009,liang2010,cui2012,shin2014,legrain2015,faglioni2016,ko2017,ackland2017,hutcheon2019,rajunatarajan2019,jerabek2022,qin2022,moriarty1982,ye1990,nogueira1999,katsnelson2000,hanfland2000,bonsignori1982,rahman1984,alouani1989,mutlu1995}.  We include only post-1980 values, as by this time the local-density approximation (LDA) \cite{kohn1965} was established as the preferred method to describe exchange and correlation (XC) in FPT.  All of the displayed results use either the LDA, or the more recent generalized-gradient approximation \cite{ziesche1998} (PBE GGA or GGAsol, typically) for XC.  Results from the literature that clearly appear to be outliers with respect to the predominant majority of results (such as $\Delta U_{\rm el}(0) \approx -11$ meV for fcc Li \cite{trickey2016}) are not included.  In each of the panels we also plot (as the horizontal lines) values (or lower limits) of $\Delta U_{\rm el}(0)$ from Table \ref{table3}. We now  discuss each metal in turn.

Our $\Delta U_{\rm el}(0)$ results for Li exhibit mixed agreement with the collection of FPT values.  First, $\Delta U_{\rm el}^{9R}(0) = -1.80 \pm 0.13$ (short-dashed violet dashed line in Fig.~\ref{Fig9}) is quite close to the majority of FPT values for the $9R$ phase, which fall within the range $-1.1$ meV to $-2.0$ meV.  Agreement for the fcc phase, however, is less satisfying overall.  Our result $\Delta U_{\rm el}^{\rm fcc}(0) = -3.60 \pm 0.23$ (green dashed line), while in good agreement with the early AE calculations (which range from $-2.9$ meV to $-4.2$ meV), is rather much larger in magnitude than PP results (which range from $-0.1$ meV to $-2.6$ meV).  The FPT results for the hcp phase do, however, lend support to our above supposition that this phase is thermodynamically unstable at all temperatures.  Notice that on average $\Delta U_{\rm el}^{\rm hcp}(0)$ values from FPT are above those for the fcc phase, and---more importantly---many of these values lie above our lower limit of $-0.99$ meV (red dashed line) for the hcp phase to be unstable.

The discrepancy between our result $\Delta U_{\rm el}^{\rm fcc}(0) = -3.60 \pm 0.23$ and FPT PP values (between $-0.1$ meV and $-2.6$ meV) might arise from several factors.  First, while it is certainly possible that PP FPT exhibits a systematic error that underestimates the magnitude of $\Delta U_{\rm el}^{\rm fcc}(0)$, we have no way of investigating this possibility; we shall thus let this possibility lie dormant.  What potential systematic issues with our calculations might account for the discrepancy?  As we noted above in discussion of Fig.~\ref{Fig5}(a), our EAM DOS and the BvK FC DOS for bcc Li show appreciable differences.  We have thus calculated values of $\Delta U_{\rm el}(0)$ using the BvK FC DOS as a baseline for differences with the cp structures. We thence find $\Delta U_{\rm el}^{\rm fcc}(0) = -3.1$ meV.  While $\Delta U_{\rm el}^{\rm fcc}(0)$ has shifted towards the FPT values, the agreement can only be regarded as marginally improved.  It is also possible that vibrational frequencies for our cp  structures are systematically off by some small factor:  as is obvious in Fig.~\ref{Fig6}, our calculated values of $\Delta U_{\rm vib}(0)$ [$= (9/8) k_B \Delta \Theta_1$] for Li are close to the upper range of all calculated values.  We have thus investigated the consequence of multiplying the frequencies of the cp structures by a factor slightly less than unity.  For example, if we multiply the frequencies by 0.975 (a 2.5\% reduction), then we find $\Delta U_{\rm el}^{\rm fcc}(0) = -2.2$ meV, which can certainly be regarded as satisfactory agreement.  In this case we also obtain $\Delta U_{\rm el}^{9R}(0) = -0.8$ meV and $\Delta U_{\rm el}^{\rm hcp}(0) > -0.04$ meV, which are also not unreasonable values in comparison with FPT.  Lastly, it is not impossible that $T_{\rm bcc}^{\rm fcc}$ is somewhat lower than the temperature $167.5 \pm 12.5$ K we identify above as equal to $T_{\rm bcc}^{\rm fcc}$.  Given that upon cooling the bcc phase does not directly convert to fcc, it may also be that case that fcc does not convert back to bcc until the temperature is somewhat above the thermodynamic transition temperature $T_{\rm bcc}^{\rm fcc}$.  If $T_{\rm bcc}^{\rm fcc}$ is indeed lower than $167.5 \pm 12.5$ K, then $\Delta U_{\rm el}^{\rm fcc}(0) = -3.60 \pm 0.23$ can only be considered a lower limit, with the actual value being somewhat closer to the FPT PP values.

What about Na?  As is evident in Fig.~\ref{Fig9}, agreement between our values for $\Delta U_{\rm el}$ and values from FPT can be regarded as satisfactory.  Our results $\Delta U_{\rm el}^{9R}(0) = -0.53 \pm 0.04$ and $\Delta U_{\rm el}^{\rm hcp}(0) = -0.69\pm 0.13$ both have magnitude less than 1 meV, as do most of the FPT values.  Given the $\sim$2 meV scatter in FPT values, a more elaborate comparison is not really possible, however.

A comparison of our results for K and Rb with FPT calculations is also quite satisfactory, in that most FPT values for $\Delta U_{\rm el}(0)$ are above our lower limits for this quantity.  As with Na, it is disappointing that FPT values vary by as much as they do, which precludes more precise comparison.

\section{Summary}
\label{SecV}

Using an EAM model specifically designed to accurately describe vibrations in bcc alkali metals, we have studied the vibrational structure of Li, Na, K, and Rb in not only the bcc configuration, but also in the close-packed (cp) fcc, hcp, and $9R$ structures.  Specifically, for all four phases we have calculated (i) dispersion curves along high-symmetry directions, (ii) vibrational densities of states, (iii) moment Debye temperatures, and (iv) the vibrational contribution $A_{\rm vib}(T)$ to the  Helmholtz energy $A(T)$.

Li and Na each undergo martensitic phase transitions from bcc to a combination of cp phases (predominantly fcc and $9R$ for Li and hcp and $9R$ for Na) at low temperatures.  Relevant to the phase transitions are three zero-temperature energies associated with each phase:  the zero-point vibrational energy $U_{\rm vib}(0)$, the electronic ground-state energy  $U_{\rm el}(0)$, and the Helmholtz energy $A(0) = U_{\rm vib}(0) + U_{\rm el}(0)$.  In conjunction with phase-transition temperatures identified from experimental work on Li and Na, our results for $A_{\rm vib}(T)$ allow us to deduce differences in $U_{\rm vib}(0)$, $U_{\rm el}(0)$, and $A(0)$ between the bcc and relevant cp phases.  For K and Rb---which do not undergo any phase transitions from bcc---we are able to infer lower bounds on differences in $U_{\rm el}(0)$.




\bibliography{AlkaliMetals}

\begin{thebibliography}{108}%
\makeatletter
\providecommand \@ifxundefined [1]{%
 \@ifx{#1\undefined}
}%
\providecommand \@ifnum [1]{%
 \ifnum #1\expandafter \@firstoftwo
 \else \expandafter \@secondoftwo
 \fi
}%
\providecommand \@ifx [1]{%
 \ifx #1\expandafter \@firstoftwo
 \else \expandafter \@secondoftwo
 \fi
}%
\providecommand \natexlab [1]{#1}%
\providecommand \enquote  [1]{``#1''}%
\providecommand \bibnamefont  [1]{#1}%
\providecommand \bibfnamefont [1]{#1}%
\providecommand \citenamefont [1]{#1}%
\providecommand \href@noop [0]{\@secondoftwo}%
\providecommand \href [0]{\begingroup \@sanitize@url \@href}%
\providecommand \@href[1]{\@@startlink{#1}\@@href}%
\providecommand \@@href[1]{\endgroup#1\@@endlink}%
\providecommand \@sanitize@url [0]{\catcode `\\12\catcode `\$12\catcode
  `\&12\catcode `\#12\catcode `\^12\catcode `\_12\catcode `\%12\relax}%
\providecommand \@@startlink[1]{}%
\providecommand \@@endlink[0]{}%
\providecommand \url  [0]{\begingroup\@sanitize@url \@url }%
\providecommand \@url [1]{\endgroup\@href {#1}{\urlprefix }}%
\providecommand \urlprefix  [0]{URL }%
\providecommand \Eprint [0]{\href }%
\providecommand \doibase [0]{http://dx.doi.org/}%
\providecommand \selectlanguage [0]{\@gobble}%
\providecommand \bibinfo  [0]{\@secondoftwo}%
\providecommand \bibfield  [0]{\@secondoftwo}%
\providecommand \translation [1]{[#1]}%
\providecommand \BibitemOpen [0]{}%
\providecommand \bibitemStop [0]{}%
\providecommand \bibitemNoStop [0]{.\EOS\space}%
\providecommand \EOS [0]{\spacefactor3000\relax}%
\providecommand \BibitemShut  [1]{\csname bibitem#1\endcsname}%
\let\auto@bib@innerbib\@empty
\bibitem [{\citenamefont {Barrett}(1947)}]{barrett1947}%
  \BibitemOpen
  \bibfield  {author} {\bibinfo {author} {\bibfnamefont {C.~S.}\ \bibnamefont
  {Barrett}},\ }\bibfield  {title} {\enquote {\bibinfo {title} {A low
  temperature transformation in lithium},}\ }\href
  {https://doi.org/10.1103/PhysRev.72.245} {\bibfield  {journal} {\bibinfo
  {journal} {Phys. Rev.}\ }\textbf {\bibinfo {volume} {72}},\ \bibinfo {pages}
  {245} (\bibinfo {year} {1947})}\BibitemShut {NoStop}%
\bibitem [{\citenamefont {Barrett}\ and\ \citenamefont
  {Trautz}(1948)}]{barrett1948}%
  \BibitemOpen
  \bibfield  {author} {\bibinfo {author} {\bibfnamefont {C.~S.}\ \bibnamefont
  {Barrett}}\ and\ \bibinfo {author} {\bibfnamefont {O.~R.}\ \bibnamefont
  {Trautz}},\ }\bibfield  {title} {\enquote {\bibinfo {title} {Low temperature
  transformations in lithium and lithium-magnesium alloys, technical
  publication \# 2346, {AIME, Institute Metals Division}},}\ }\href@noop {}
  {\bibfield  {journal} {\bibinfo  {journal} {Metals Technology}\ }\textbf
  {\bibinfo {volume} {15}},\ \bibinfo {pages} {579--605} (\bibinfo {year}
  {1948})}\BibitemShut {NoStop}%
\bibitem [{\citenamefont {Hovi}\ \emph {et~al.}(1966)\citenamefont {Hovi},
  \citenamefont {M{\"a}ntysalo},\ and\ \citenamefont {Tiusanen}}]{hovi1966}%
  \BibitemOpen
  \bibfield  {author} {\bibinfo {author} {\bibfnamefont {V.}~\bibnamefont
  {Hovi}}, \bibinfo {author} {\bibfnamefont {E.}~\bibnamefont {M{\"a}ntysalo}},
  \ and\ \bibinfo {author} {\bibfnamefont {K.}~\bibnamefont {Tiusanen}},\
  }\bibfield  {title} {\enquote {\bibinfo {title} {Determination of the
  martensitic critical temperature of the lithium single crystal by using the
  {L}aue method},}\ }\href {https://doi.org/10.1016/0001-6160(66)90280-X}
  {\bibfield  {journal} {\bibinfo  {journal} {Acta Metallurgica}\ }\textbf
  {\bibinfo {volume} {14}},\ \bibinfo {pages} {67--69} (\bibinfo {year}
  {1966})}\BibitemShut {NoStop}%
\bibitem [{\citenamefont {McCarthy}\ \emph {et~al.}(1980)\citenamefont
  {McCarthy}, \citenamefont {Tompson},\ and\ \citenamefont
  {Werner}}]{mccarthy1980}%
  \BibitemOpen
  \bibfield  {author} {\bibinfo {author} {\bibfnamefont {C.~M.}\ \bibnamefont
  {McCarthy}}, \bibinfo {author} {\bibfnamefont {C.~W.}\ \bibnamefont
  {Tompson}}, \ and\ \bibinfo {author} {\bibfnamefont {S~A.}\ \bibnamefont
  {Werner}},\ }\bibfield  {title} {\enquote {\bibinfo {title} {Anharmonicity
  and the low-temperature phase in lithium metal},}\ }\href
  {https://doi.org/10.1103/PhysRevB.22.574} {\bibfield  {journal} {\bibinfo
  {journal} {Phys. Rev. B}\ }\textbf {\bibinfo {volume} {22}},\ \bibinfo
  {pages} {574--580} (\bibinfo {year} {1980})}\BibitemShut {NoStop}%
\bibitem [{\citenamefont {Berliner}\ and\ \citenamefont
  {Werner}(1986{\natexlab{a}})}]{berliner1986A}%
  \BibitemOpen
  \bibfield  {author} {\bibinfo {author} {\bibfnamefont {R.}~\bibnamefont
  {Berliner}}\ and\ \bibinfo {author} {\bibfnamefont {S.~A.}\ \bibnamefont
  {Werner}},\ }\bibfield  {title} {\enquote {\bibinfo {title} {The structure of
  the low temperature phase of lithium metal},}\ }\href
  {https://doi.org/10.1016/S0378-4363(86)80123-1} {\bibfield  {journal}
  {\bibinfo  {journal} {Physica B+C}\ }\textbf {\bibinfo {volume} {136}},\
  \bibinfo {pages} {481--484} (\bibinfo {year}
  {1986}{\natexlab{a}})}\BibitemShut {NoStop}%
\bibitem [{\citenamefont {Smith}(1987)}]{smith1987}%
  \BibitemOpen
  \bibfield  {author} {\bibinfo {author} {\bibfnamefont {H.~G.}\ \bibnamefont
  {Smith}},\ }\bibfield  {title} {\enquote {\bibinfo {title} {Martensitic phase
  transformation of single-crystal lithium from bcc to a 9{$R$}-related
  structure},}\ }\href {https://doi.org/10.1103/PhysRevLett.58.1228} {\bibfield
   {journal} {\bibinfo  {journal} {Phys. Rev. Lett.}\ }\textbf {\bibinfo
  {volume} {58}},\ \bibinfo {pages} {1228--1231} (\bibinfo {year}
  {1987})}\BibitemShut {NoStop}%
\bibitem [{\citenamefont {Schwarz}\ and\ \citenamefont
  {Blaschko}(1990)}]{schwarz1990}%
  \BibitemOpen
  \bibfield  {author} {\bibinfo {author} {\bibfnamefont {W.}~\bibnamefont
  {Schwarz}}\ and\ \bibinfo {author} {\bibfnamefont {O.}~\bibnamefont
  {Blaschko}},\ }\bibfield  {title} {\enquote {\bibinfo {title} {Polytype
  structures of lithium at low temperatures},}\ }\href
  {https://doi.org/10.1103/PhysRevLett.65.3144} {\bibfield  {journal} {\bibinfo
   {journal} {Phys. Rev. Lett.}\ }\textbf {\bibinfo {volume} {65}},\ \bibinfo
  {pages} {3144--3147} (\bibinfo {year} {1990})}\BibitemShut {NoStop}%
\bibitem [{\citenamefont {Schwarz}\ \emph {et~al.}(1991)\citenamefont
  {Schwarz}, \citenamefont {Blaschko},\ and\ \citenamefont
  {Gorgas}}]{schwarz1991}%
  \BibitemOpen
  \bibfield  {author} {\bibinfo {author} {\bibfnamefont {W.}~\bibnamefont
  {Schwarz}}, \bibinfo {author} {\bibfnamefont {O.}~\bibnamefont {Blaschko}}, \
  and\ \bibinfo {author} {\bibfnamefont {I.}~\bibnamefont {Gorgas}},\
  }\bibfield  {title} {\enquote {\bibinfo {title} {bcc instability of lithium
  at low temperatures},}\ }\href {https://doi.org/10.1103/PhysRevB.44.6785}
  {\bibfield  {journal} {\bibinfo  {journal} {Phys. Rev.}\ }\textbf {\bibinfo
  {volume} {44}},\ \bibinfo {pages} {6785} (\bibinfo {year}
  {1991})}\BibitemShut {NoStop}%
\bibitem [{\citenamefont {Berliner}\ and\ \citenamefont
  {Werner}(1986{\natexlab{b}})}]{berliner1986B}%
  \BibitemOpen
  \bibfield  {author} {\bibinfo {author} {\bibfnamefont {R.}~\bibnamefont
  {Berliner}}\ and\ \bibinfo {author} {\bibfnamefont {S.~A.}\ \bibnamefont
  {Werner}},\ }\bibfield  {title} {\enquote {\bibinfo {title} {Effect of
  stacking faults on diffraction: {T}he structure of lithium metal},}\ }\href
  {https://doi.org/10.1103/PhysRevB.34.3586} {\bibfield  {journal} {\bibinfo
  {journal} {Phys. Rev. B}\ }\textbf {\bibinfo {volume} {34}},\ \bibinfo
  {pages} {3586--3603} (\bibinfo {year} {1986}{\natexlab{b}})}\BibitemShut
  {NoStop}%
\bibitem [{\citenamefont {Berliner}\ \emph {et~al.}(1989)\citenamefont
  {Berliner}, \citenamefont {Fajen}, \citenamefont {Smith},\ and\ \citenamefont
  {Hitterman}}]{berliner1989}%
  \BibitemOpen
  \bibfield  {author} {\bibinfo {author} {\bibfnamefont {R.}~\bibnamefont
  {Berliner}}, \bibinfo {author} {\bibfnamefont {O.}~\bibnamefont {Fajen}},
  \bibinfo {author} {\bibfnamefont {H.~G.}\ \bibnamefont {Smith}}, \ and\
  \bibinfo {author} {\bibfnamefont {R.~L.}\ \bibnamefont {Hitterman}},\
  }\bibfield  {title} {\enquote {\bibinfo {title} {Neutron powder-diffraction
  studies of lithium, sodium, and potassium metal},}\ }\href
  {https://doi.org/10.1103/PhysRevB.40.12086} {\bibfield  {journal} {\bibinfo
  {journal} {Phys. Rev. B}\ }\textbf {\bibinfo {volume} {40}},\ \bibinfo
  {pages} {12086--12096} (\bibinfo {year} {1989})}\BibitemShut {NoStop}%
\bibitem [{\citenamefont {Smith}\ \emph {et~al.}(1994)\citenamefont {Smith},
  \citenamefont {Berliner},\ and\ \citenamefont {Trivisonno}}]{smith1994}%
  \BibitemOpen
  \bibfield  {author} {\bibinfo {author} {\bibfnamefont {H.~G.}\ \bibnamefont
  {Smith}}, \bibinfo {author} {\bibfnamefont {R.}~\bibnamefont {Berliner}}, \
  and\ \bibinfo {author} {\bibfnamefont {J.}~\bibnamefont {Trivisonno}},\
  }\bibfield  {title} {\enquote {\bibinfo {title} {Absence of precursor effects
  above the martensitic transformation in a virgin crystal of {Li} metal},}\
  }\href {https://doi.org/10.1103/PhysRevB.49.8547} {\bibfield  {journal}
  {\bibinfo  {journal} {Phys. Rev. B}\ }\textbf {\bibinfo {volume} {49}},\
  \bibinfo {pages} {8547--8551} (\bibinfo {year} {1994})}\BibitemShut {NoStop}%
\bibitem [{\citenamefont {Maier}\ \emph {et~al.}(1995)\citenamefont {Maier},
  \citenamefont {Blaschko},\ and\ \citenamefont {Pichl}}]{maier1995}%
  \BibitemOpen
  \bibfield  {author} {\bibinfo {author} {\bibfnamefont {Ch.}\ \bibnamefont
  {Maier}}, \bibinfo {author} {\bibfnamefont {O.}~\bibnamefont {Blaschko}}, \
  and\ \bibinfo {author} {\bibfnamefont {W.}~\bibnamefont {Pichl}},\ }\bibfield
   {title} {\enquote {\bibinfo {title} {Influence of uniaxial deformation on
  the phase transformation in lithium},}\ }\href
  {https://doi.org/10.1103/PhysRevB.52.9283} {\bibfield  {journal} {\bibinfo
  {journal} {Phys. Rev. B}\ }\textbf {\bibinfo {volume} {52}},\ \bibinfo
  {pages} {9283--9290} (\bibinfo {year} {1995})}\BibitemShut {NoStop}%
\bibitem [{\citenamefont {Ackland}\ \emph {et~al.}(2017)\citenamefont
  {Ackland}, \citenamefont {Dunuwille}, \citenamefont {Martinez-Canales},
  \citenamefont {Loa}, \citenamefont {Zhang}, \citenamefont {Sinogeikin},
  \citenamefont {Cai},\ and\ \citenamefont {Deemyad}}]{ackland2017}%
  \BibitemOpen
  \bibfield  {author} {\bibinfo {author} {\bibfnamefont {G.~J.}\ \bibnamefont
  {Ackland}}, \bibinfo {author} {\bibfnamefont {M.}~\bibnamefont {Dunuwille}},
  \bibinfo {author} {\bibfnamefont {M.}~\bibnamefont {Martinez-Canales}},
  \bibinfo {author} {\bibfnamefont {I.}~\bibnamefont {Loa}}, \bibinfo {author}
  {\bibfnamefont {R.}~\bibnamefont {Zhang}}, \bibinfo {author} {\bibfnamefont
  {S.}~\bibnamefont {Sinogeikin}}, \bibinfo {author} {\bibfnamefont
  {W.}~\bibnamefont {Cai}}, \ and\ \bibinfo {author} {\bibfnamefont
  {S.}~\bibnamefont {Deemyad}},\ }\bibfield  {title} {\enquote {\bibinfo
  {title} {Quantum and isotope effects in lithium metal},}\ }\href
  {https://doi.org/10.1126/science.aal4886} {\bibfield  {journal} {\bibinfo
  {journal} {Science}\ }\textbf {\bibinfo {volume} {356}},\ \bibinfo {pages}
  {1254--1259} (\bibinfo {year} {2017})}\BibitemShut {NoStop}%
\bibitem [{\citenamefont {Owen}\ and\ \citenamefont
  {Williams}(1954)}]{owen1954}%
  \BibitemOpen
  \bibfield  {author} {\bibinfo {author} {\bibfnamefont {E.~A.}\ \bibnamefont
  {Owen}}\ and\ \bibinfo {author} {\bibfnamefont {G.~I.}\ \bibnamefont
  {Williams}},\ }\bibfield  {title} {\enquote {\bibinfo {title} {X-ray
  measurements on lithium at low temperatures},}\ }\href
  {https://doi.org/10.1088/0370-1298/67/10/306} {\bibfield  {journal} {\bibinfo
   {journal} {Proceedings of the Physical Society. Section A}\ }\textbf
  {\bibinfo {volume} {67}},\ \bibinfo {pages} {895--900} (\bibinfo {year}
  {1954})}\BibitemShut {NoStop}%
\bibitem [{\citenamefont {Ernst}\ \emph {et~al.}(1986)\citenamefont {Ernst},
  \citenamefont {Artner}, \citenamefont {Blaschko},\ and\ \citenamefont
  {Krexner}}]{ernst1986}%
  \BibitemOpen
  \bibfield  {author} {\bibinfo {author} {\bibfnamefont {G.}~\bibnamefont
  {Ernst}}, \bibinfo {author} {\bibfnamefont {C.}~\bibnamefont {Artner}},
  \bibinfo {author} {\bibfnamefont {O.}~\bibnamefont {Blaschko}}, \ and\
  \bibinfo {author} {\bibfnamefont {G.}~\bibnamefont {Krexner}},\ }\bibfield
  {title} {\enquote {\bibinfo {title} {Low-temperature martensitic phase
  transition of bcc lithium},}\ }\href
  {https://doi.org/10.1103/PhysRevB.33.6465} {\bibfield  {journal} {\bibinfo
  {journal} {Phys. Rev. B}\ }\textbf {\bibinfo {volume} {33}},\ \bibinfo
  {pages} {6465--6469} (\bibinfo {year} {1986})}\BibitemShut {NoStop}%
\bibitem [{\citenamefont {Barrett}(1956)}]{BarrettAC1956}%
  \BibitemOpen
  \bibfield  {author} {\bibinfo {author} {\bibfnamefont {C.~S.}\ \bibnamefont
  {Barrett}},\ }\bibfield  {title} {\enquote {\bibinfo {title} {{X-ray study of
  the alkali metals at low temperatures}},}\ }\href
  {https://doi.org/10.1107/S0365110X56001790} {\bibfield  {journal} {\bibinfo
  {journal} {Acta Crystallographica}\ }\textbf {\bibinfo {volume} {9}},\
  \bibinfo {pages} {671} (\bibinfo {year} {1956})}\BibitemShut {NoStop}%
\bibitem [{\citenamefont {Stedman}(1976)}]{stedman1976}%
  \BibitemOpen
  \bibfield  {author} {\bibinfo {author} {\bibfnamefont {R.}~\bibnamefont
  {Stedman}},\ }\bibfield  {title} {\enquote {\bibinfo {title} {Observations on
  the martensitic transition in sodium},}\ }\href
  {https://doi.org/10.1088/0305-4608/6/12/010} {\bibfield  {journal} {\bibinfo
  {journal} {Journal of Physics F: Metal Physics}\ }\textbf {\bibinfo {volume}
  {6}},\ \bibinfo {pages} {2239--2246} (\bibinfo {year} {1976})}\BibitemShut
  {NoStop}%
\bibitem [{\citenamefont {Schwarz}\ \emph {et~al.}(1992)\citenamefont
  {Schwarz}, \citenamefont {Blaschko},\ and\ \citenamefont
  {Gorgas}}]{schwarz1992}%
  \BibitemOpen
  \bibfield  {author} {\bibinfo {author} {\bibfnamefont {W.}~\bibnamefont
  {Schwarz}}, \bibinfo {author} {\bibfnamefont {O.}~\bibnamefont {Blaschko}}, \
  and\ \bibinfo {author} {\bibfnamefont {I.}~\bibnamefont {Gorgas}},\
  }\bibfield  {title} {\enquote {\bibinfo {title} {Diffuse-neutron-scattering
  investigation of the low-temperature phases of sodium},}\ }\href
  {https://doi.org/10.1103/PhysRevB.46.14448} {\bibfield  {journal} {\bibinfo
  {journal} {Phys. Rev. B}\ }\textbf {\bibinfo {volume} {46}},\ \bibinfo
  {pages} {14448--14452} (\bibinfo {year} {1992})}\BibitemShut {NoStop}%
\bibitem [{\citenamefont {Berliner}\ \emph {et~al.}(1992)\citenamefont
  {Berliner}, \citenamefont {Smith}, \citenamefont {Copley},\ and\
  \citenamefont {Trivisonno}}]{berliner1992}%
  \BibitemOpen
  \bibfield  {author} {\bibinfo {author} {\bibfnamefont {R.}~\bibnamefont
  {Berliner}}, \bibinfo {author} {\bibfnamefont {H.~G.}\ \bibnamefont {Smith}},
  \bibinfo {author} {\bibfnamefont {J.~R.~D.}\ \bibnamefont {Copley}}, \ and\
  \bibinfo {author} {\bibfnamefont {J.}~\bibnamefont {Trivisonno}},\ }\bibfield
   {title} {\enquote {\bibinfo {title} {Structures of sodium metal},}\ }\href
  {https://doi.org/10.1103/PhysRevB.46.14436} {\bibfield  {journal} {\bibinfo
  {journal} {Phys. Rev. B}\ }\textbf {\bibinfo {volume} {46}},\ \bibinfo
  {pages} {14436--14447} (\bibinfo {year} {1992})}\BibitemShut {NoStop}%
\bibitem [{\citenamefont {Abe}\ \emph {et~al.}(1994)\citenamefont {Abe},
  \citenamefont {Ohshima}, \citenamefont {Suzuki}, \citenamefont {Hoshino},\
  and\ \citenamefont {Kakurai}}]{abe1994}%
  \BibitemOpen
  \bibfield  {author} {\bibinfo {author} {\bibfnamefont {H.}~\bibnamefont
  {Abe}}, \bibinfo {author} {\bibfnamefont {K.}~\bibnamefont {Ohshima}},
  \bibinfo {author} {\bibfnamefont {T.}~\bibnamefont {Suzuki}}, \bibinfo
  {author} {\bibfnamefont {S.}~\bibnamefont {Hoshino}}, \ and\ \bibinfo
  {author} {\bibfnamefont {K.}~\bibnamefont {Kakurai}},\ }\bibfield  {title}
  {\enquote {\bibinfo {title} {Neutron-scattering study of metallic sodium},}\
  }\href {https://doi.org/10.1103/PhysRevB.49.3739} {\bibfield  {journal}
  {\bibinfo  {journal} {Phys. Rev. B}\ }\textbf {\bibinfo {volume} {49}},\
  \bibinfo {pages} {3739--3745} (\bibinfo {year} {1994})}\BibitemShut {NoStop}%
\bibitem [{\citenamefont {Abe}\ \emph {et~al.}(1997)\citenamefont {Abe},
  \citenamefont {Matsuo}, \citenamefont {Ishibashi}, \citenamefont {Ohshima},
  \citenamefont {Imai},\ and\ \citenamefont {Kakurai}}]{abe1997}%
  \BibitemOpen
  \bibfield  {author} {\bibinfo {author} {\bibfnamefont {H.}~\bibnamefont
  {Abe}}, \bibinfo {author} {\bibfnamefont {R.~J.}\ \bibnamefont {Matsuo}},
  \bibinfo {author} {\bibfnamefont {M.}~\bibnamefont {Ishibashi}}, \bibinfo
  {author} {\bibfnamefont {K.-i.}\ \bibnamefont {Ohshima}}, \bibinfo {author}
  {\bibfnamefont {M.}~\bibnamefont {Imai}}, \ and\ \bibinfo {author}
  {\bibfnamefont {K.}~\bibnamefont {Kakurai}},\ }\bibfield  {title} {\enquote
  {\bibinfo {title} {Small-angle neutron scattering study of metallic sodium on
  phase transition},}\ }\href {https://doi.org/10.1143/jpsj.66.1860} {\bibfield
   {journal} {\bibinfo  {journal} {Journal of the Physical Society of Japan}\
  }\textbf {\bibinfo {volume} {66}},\ \bibinfo {pages} {1860--1863} (\bibinfo
  {year} {1997})}\BibitemShut {NoStop}%
\bibitem [{\citenamefont {Blaschko}\ and\ \citenamefont
  {Krexner}(1984)}]{blaschko1984}%
  \BibitemOpen
  \bibfield  {author} {\bibinfo {author} {\bibfnamefont {O.}~\bibnamefont
  {Blaschko}}\ and\ \bibinfo {author} {\bibfnamefont {G.}~\bibnamefont
  {Krexner}},\ }\bibfield  {title} {\enquote {\bibinfo {title} {Observation of
  static and dynamic pretransitional phenomena near the martensitic bcc-hcp
  transformation in sodium},}\ }\href
  {https://doi.org/10.1103/PhysRevB.30.1667} {\bibfield  {journal} {\bibinfo
  {journal} {Phys. Rev. B}\ }\textbf {\bibinfo {volume} {30}},\ \bibinfo
  {pages} {1667--1670} (\bibinfo {year} {1984})}\BibitemShut {NoStop}%
\bibitem [{\citenamefont {Szente}\ and\ \citenamefont
  {Trivisonno}(1988)}]{szente1988}%
  \BibitemOpen
  \bibfield  {author} {\bibinfo {author} {\bibfnamefont {J.}~\bibnamefont
  {Szente}}\ and\ \bibinfo {author} {\bibfnamefont {J.}~\bibnamefont
  {Trivisonno}},\ }\bibfield  {title} {\enquote {\bibinfo {title} {Ultrasonic
  study of the martensitic phase transformation in sodium},}\ }\href
  {https://doi.org/10.1103/PhysRevB.37.8447} {\bibfield  {journal} {\bibinfo
  {journal} {Phys. Rev. B}\ }\textbf {\bibinfo {volume} {37}},\ \bibinfo
  {pages} {8447--8450} (\bibinfo {year} {1988})}\BibitemShut {NoStop}%
\bibitem [{\citenamefont {Overhauser}(1984)}]{overhauser1984}%
  \BibitemOpen
  \bibfield  {author} {\bibinfo {author} {\bibfnamefont {A.~W.}\ \bibnamefont
  {Overhauser}},\ }\bibfield  {title} {\enquote {\bibinfo {title} {Crystal
  structure of lithium at 4.2 {K}},}\ }\href
  {https://doi.org/10.1103/PhysRevLett.53.64} {\bibfield  {journal} {\bibinfo
  {journal} {Phys. Rev. Lett.}\ }\textbf {\bibinfo {volume} {53}},\ \bibinfo
  {pages} {64--65} (\bibinfo {year} {1984})}\BibitemShut {NoStop}%
\bibitem [{\citenamefont {Kubinski}\ and\ \citenamefont
  {Trivisonno}(1993)}]{kubinski1993}%
  \BibitemOpen
  \bibfield  {author} {\bibinfo {author} {\bibfnamefont {D.}~\bibnamefont
  {Kubinski}}\ and\ \bibinfo {author} {\bibfnamefont {J.}~\bibnamefont
  {Trivisonno}},\ }\bibfield  {title} {\enquote {\bibinfo {title} {Absence of a
  martensitic phase transformation (even an embryonic phase) in potassium as
  determined by an ultrasound study},}\ }\href
  {https://doi.org/10.1103/PhysRevB.47.1069} {\bibfield  {journal} {\bibinfo
  {journal} {Phys. Rev. B}\ }\textbf {\bibinfo {volume} {47}},\ \bibinfo
  {pages} {1069--1072} (\bibinfo {year} {1993})}\BibitemShut {NoStop}%
\bibitem [{\citenamefont {Schneider}\ and\ \citenamefont
  {Stoll}(1970)}]{schneider1970}%
  \BibitemOpen
  \bibfield  {author} {\bibinfo {author} {\bibfnamefont {T.}~\bibnamefont
  {Schneider}}\ and\ \bibinfo {author} {\bibfnamefont {E.}~\bibnamefont
  {Stoll}},\ }\bibfield  {title} {\enquote {\bibinfo {title} {On the nature of
  the martensitic phase transformation in lithium},}\ }\href
  {https://doi.org/10.1016/0038-1098(70)90384-4} {\bibfield  {journal}
  {\bibinfo  {journal} {Solid State Commun.}\ }\textbf {\bibinfo {volume}
  {8}},\ \bibinfo {pages} {1729--1731} (\bibinfo {year} {1970})}\BibitemShut
  {NoStop}%
\bibitem [{\citenamefont {Bajpai}\ \emph {et~al.}(1975)\citenamefont {Bajpai},
  \citenamefont {Ono}, \citenamefont {Ohno},\ and\ \citenamefont
  {Toya}}]{bajpai1975}%
  \BibitemOpen
  \bibfield  {author} {\bibinfo {author} {\bibfnamefont {R.~P.}\ \bibnamefont
  {Bajpai}}, \bibinfo {author} {\bibfnamefont {M.}~\bibnamefont {Ono}},
  \bibinfo {author} {\bibfnamefont {Y.}~\bibnamefont {Ohno}}, \ and\ \bibinfo
  {author} {\bibfnamefont {T.}~\bibnamefont {Toya}},\ }\bibfield  {title}
  {\enquote {\bibinfo {title} {Self-consistent theory of the martensitic phase
  transformation in metallic lithium},}\ }\href
  {https://doi.org/10.1103/PhysRevB.12.2194} {\bibfield  {journal} {\bibinfo
  {journal} {Phys. Rev. B}\ }\textbf {\bibinfo {volume} {12}},\ \bibinfo
  {pages} {2194--2211} (\bibinfo {year} {1975})}\BibitemShut {NoStop}%
\bibitem [{\citenamefont {Hutcheon}\ and\ \citenamefont
  {Needs}(2019)}]{hutcheon2019}%
  \BibitemOpen
  \bibfield  {author} {\bibinfo {author} {\bibfnamefont {M.}~\bibnamefont
  {Hutcheon}}\ and\ \bibinfo {author} {\bibfnamefont {R.}~\bibnamefont
  {Needs}},\ }\bibfield  {title} {\enquote {\bibinfo {title} {Structural and
  vibrational properties of lithium under ambient conditions within density
  functional theory},}\ }\href {https://doi.org/10.1103/PhysRevB.99.014111}
  {\bibfield  {journal} {\bibinfo  {journal} {Phys. Rev. B}\ }\textbf {\bibinfo
  {volume} {99}},\ \bibinfo {pages} {014111} (\bibinfo {year}
  {2019})}\BibitemShut {NoStop}%
\bibitem [{\citenamefont {Staikov}\ \emph {et~al.}(1997)\citenamefont
  {Staikov}, \citenamefont {Kara},\ and\ \citenamefont {Rahman}}]{staikov1997}%
  \BibitemOpen
  \bibfield  {author} {\bibinfo {author} {\bibfnamefont {P.}~\bibnamefont
  {Staikov}}, \bibinfo {author} {\bibfnamefont {A.}~\bibnamefont {Kara}}, \
  and\ \bibinfo {author} {\bibfnamefont {T.~S.}\ \bibnamefont {Rahman}},\
  }\bibfield  {title} {\enquote {\bibinfo {title} {First-principles studies of
  the thermodynamic properties of bulk {Li}},}\ }\href
  {https://doi.org/10.1088/0953-8984/9/10/004} {\bibfield  {journal} {\bibinfo
  {journal} {J. Phys.: Condens. Matter}\ }\textbf {\bibinfo {volume} {9}},\
  \bibinfo {pages} {2135--2148} (\bibinfo {year} {1997})}\BibitemShut {NoStop}%
\bibitem [{\citenamefont {Yao}\ \emph {et~al.}(2009)\citenamefont {Yao},
  \citenamefont {Tse}, \citenamefont {Tanaka}, \citenamefont {Marsiglio},\ and\
  \citenamefont {Ma}}]{yao2009}%
  \BibitemOpen
  \bibfield  {author} {\bibinfo {author} {\bibfnamefont {Y.}~\bibnamefont
  {Yao}}, \bibinfo {author} {\bibfnamefont {J.~S.}\ \bibnamefont {Tse}},
  \bibinfo {author} {\bibfnamefont {K.}~\bibnamefont {Tanaka}}, \bibinfo
  {author} {\bibfnamefont {F.}~\bibnamefont {Marsiglio}}, \ and\ \bibinfo
  {author} {\bibfnamefont {Y.}~\bibnamefont {Ma}},\ }\bibfield  {title}
  {\enquote {\bibinfo {title} {Superconductivity in lithium under high pressure
  investigated with density functional and {E}liashberg theory},}\ }\href
  {https://doi.org/10.1103/PhysRevB.79.054524} {\bibfield  {journal} {\bibinfo
  {journal} {Phys. Rev. B}\ }\textbf {\bibinfo {volume} {79}},\ \bibinfo
  {pages} {054524} (\bibinfo {year} {2009})}\BibitemShut {NoStop}%
\bibitem [{\citenamefont {Liu}\ \emph {et~al.}(1999)\citenamefont {Liu},
  \citenamefont {Quong}, \citenamefont {Freericks}, \citenamefont {Nicol},\
  and\ \citenamefont {Jones}}]{Liu1999}%
  \BibitemOpen
  \bibfield  {author} {\bibinfo {author} {\bibfnamefont {A.~Y.}\ \bibnamefont
  {Liu}}, \bibinfo {author} {\bibfnamefont {A.~A.}\ \bibnamefont {Quong}},
  \bibinfo {author} {\bibfnamefont {J.~K.}\ \bibnamefont {Freericks}}, \bibinfo
  {author} {\bibfnamefont {E.~J.}\ \bibnamefont {Nicol}}, \ and\ \bibinfo
  {author} {\bibfnamefont {E.~C.}\ \bibnamefont {Jones}},\ }\bibfield  {title}
  {\enquote {\bibinfo {title} {Structural phase stability and electron-phonon
  coupling in lithium},}\ }\href {https://doi.org/10.1103/PhysRevB.59.4028}
  {\bibfield  {journal} {\bibinfo  {journal} {Phys. Rev. B}\ }\textbf {\bibinfo
  {volume} {59}},\ \bibinfo {pages} {4028--4035} (\bibinfo {year}
  {1999})}\BibitemShut {NoStop}%
\bibitem [{\citenamefont {Pynn}\ and\ \citenamefont
  {Ebbsj{\"o}}(1971)}]{pynn1971}%
  \BibitemOpen
  \bibfield  {author} {\bibinfo {author} {\bibfnamefont {R.}~\bibnamefont
  {Pynn}}\ and\ \bibinfo {author} {\bibfnamefont {I.}~\bibnamefont
  {Ebbsj{\"o}}},\ }\bibfield  {title} {\enquote {\bibinfo {title} {A study of
  the effect of temperature on the stable crystal structures of potassium and
  sodium},}\ }\href {https://doi.org/10.1088/0305-4608/1/5/328} {\bibfield
  {journal} {\bibinfo  {journal} {Journal of Physics F: Metal Physics}\
  }\textbf {\bibinfo {volume} {1}},\ \bibinfo {pages} {744--752} (\bibinfo
  {year} {1971})}\BibitemShut {NoStop}%
\bibitem [{\citenamefont {Straub}\ and\ \citenamefont
  {Wallace}(1971)}]{straub1971}%
  \BibitemOpen
  \bibfield  {author} {\bibinfo {author} {\bibfnamefont {G.~K.}\ \bibnamefont
  {Straub}}\ and\ \bibinfo {author} {\bibfnamefont {D.~C.}\ \bibnamefont
  {Wallace}},\ }\bibfield  {title} {\enquote {\bibinfo {title} {Study of the
  martensitic phase transition in sodium},}\ }\href
  {https://doi.org/10.1103/PhysRevB.3.1234} {\bibfield  {journal} {\bibinfo
  {journal} {Phys. Rev. B}\ }\textbf {\bibinfo {volume} {3}},\ \bibinfo {pages}
  {1234--1239} (\bibinfo {year} {1971})}\BibitemShut {NoStop}%
\bibitem [{\citenamefont {Boettger}\ and\ \citenamefont
  {Trickey}(1985)}]{BoettgerPRB1985}%
  \BibitemOpen
  \bibfield  {author} {\bibinfo {author} {\bibfnamefont {J.~C.}\ \bibnamefont
  {Boettger}}\ and\ \bibinfo {author} {\bibfnamefont {S.~B.}\ \bibnamefont
  {Trickey}},\ }\bibfield  {title} {\enquote {\bibinfo {title} {Equation of
  state and properties of lithium},}\ }\href
  {https://doi.org/10.1103/PhysRevB.32.3391} {\bibfield  {journal} {\bibinfo
  {journal} {Phys. Rev.}\ }\textbf {\bibinfo {volume} {32}},\ \bibinfo {pages}
  {3391} (\bibinfo {year} {1985})}\BibitemShut {NoStop}%
\bibitem [{\citenamefont {Krashaninin}\ and\ \citenamefont
  {Shunyaev}(1993)}]{krashaninin1993}%
  \BibitemOpen
  \bibfield  {author} {\bibinfo {author} {\bibfnamefont {V.~A.}\ \bibnamefont
  {Krashaninin}}\ and\ \bibinfo {author} {\bibfnamefont {K.~Y.}\ \bibnamefont
  {Shunyaev}},\ }\bibfield  {title} {\enquote {\bibinfo {title} {Simple metal
  binding energy in the pseudopotential method from first principles},}\ }\href
  {https://doi.org/10.1002/pssb.2221780231} {\bibfield  {journal} {\bibinfo
  {journal} {Physica Status Solidi (b)}\ }\textbf {\bibinfo {volume} {178}},\
  \bibinfo {pages} {K61--K65} (\bibinfo {year} {1993})}\BibitemShut {NoStop}%
\bibitem [{\citenamefont {Trickey}(2016)}]{trickey2016}%
  \BibitemOpen
  \bibfield  {author} {\bibinfo {author} {\bibfnamefont {S.~B.}\ \bibnamefont
  {Trickey}},\ }\bibfield  {title} {\enquote {\bibinfo {title} {Unexpected cold
  curve sensitivity to {GGA} exchange form},}\ }\href
  {https://doi.org/10.1007/s00214-016-1956-z} {\bibfield  {journal} {\bibinfo
  {journal} {Theoretical Chemistry Accounts}\ }\textbf {\bibinfo {volume}
  {135}},\ \bibinfo {pages} {219} (\bibinfo {year} {2016})}\BibitemShut
  {NoStop}%
\bibitem [{\citenamefont {Almeida}\ \emph {et~al.}(2003)\citenamefont
  {Almeida}, \citenamefont {Fiolhais},\ and\ \citenamefont
  {Caus{\`a}}}]{almeida2003}%
  \BibitemOpen
  \bibfield  {author} {\bibinfo {author} {\bibfnamefont {L.~M.}\ \bibnamefont
  {Almeida}}, \bibinfo {author} {\bibfnamefont {C.}~\bibnamefont {Fiolhais}}, \
  and\ \bibinfo {author} {\bibfnamefont {M.}~\bibnamefont {Caus{\`a}}},\
  }\bibfield  {title} {\enquote {\bibinfo {title} {Properties of simple metals
  beyond the local density approximation of density functional theory},}\
  }\href {https://doi.org/10.1002/qua.10399} {\bibfield  {journal} {\bibinfo
  {journal} {International Journal of Quantum Chemistry}\ }\textbf {\bibinfo
  {volume} {91}},\ \bibinfo {pages} {224--229} (\bibinfo {year}
  {2003})}\BibitemShut {NoStop}%
\bibitem [{\citenamefont {Hanfland}\ \emph {et~al.}(2000)\citenamefont
  {Hanfland}, \citenamefont {Syassen}, \citenamefont {Christensen},\ and\
  \citenamefont {Novikov}}]{hanfland2000}%
  \BibitemOpen
  \bibfield  {author} {\bibinfo {author} {\bibfnamefont {M.}~\bibnamefont
  {Hanfland}}, \bibinfo {author} {\bibfnamefont {K.}~\bibnamefont {Syassen}},
  \bibinfo {author} {\bibfnamefont {N.~E.}\ \bibnamefont {Christensen}}, \ and\
  \bibinfo {author} {\bibfnamefont {D.~L.}\ \bibnamefont {Novikov}},\
  }\bibfield  {title} {\enquote {\bibinfo {title} {New high-pressure phases of
  lithium},}\ }\href {https://doi.org/10.1038/35041515} {\bibfield  {journal}
  {\bibinfo  {journal} {Nature}\ }\textbf {\bibinfo {volume} {408}},\ \bibinfo
  {pages} {174--178} (\bibinfo {year} {2000})}\BibitemShut {NoStop}%
\bibitem [{\citenamefont {Jerabek}\ \emph {et~al.}(2022)\citenamefont
  {Jerabek}, \citenamefont {Burrows},\ and\ \citenamefont
  {Schwerdtfeger}}]{jerabek2022}%
  \BibitemOpen
  \bibfield  {author} {\bibinfo {author} {\bibfnamefont {P.}~\bibnamefont
  {Jerabek}}, \bibinfo {author} {\bibfnamefont {A.}~\bibnamefont {Burrows}}, \
  and\ \bibinfo {author} {\bibfnamefont {P.}~\bibnamefont {Schwerdtfeger}},\
  }\bibfield  {title} {\enquote {\bibinfo {title} {Solving a problem with a
  single parameter: a smooth bcc to fcc phase transition for metallic
  lithium},}\ }\href {https://doi.org/10.1039/D2CC04928G} {\bibfield  {journal}
  {\bibinfo  {journal} {Chemical Commun.}\ }\textbf {\bibinfo {volume} {58}},\
  \bibinfo {pages} {13369--13372} (\bibinfo {year} {2022})}\BibitemShut
  {NoStop}%
\bibitem [{\citenamefont {Wilson}\ and\ \citenamefont
  {Riffe}(2012)}]{Wilson2012}%
  \BibitemOpen
  \bibfield  {author} {\bibinfo {author} {\bibfnamefont {R.~B.}\ \bibnamefont
  {Wilson}}\ and\ \bibinfo {author} {\bibfnamefont {D.~M.}\ \bibnamefont
  {Riffe}},\ }\bibfield  {title} {\enquote {\bibinfo {title} {An
  embedded-atom-method model for alkali-metal vibrations},}\ }\href
  {https://doi.org/10.1088/0953-8984/24/33/335401} {\bibfield  {journal}
  {\bibinfo  {journal} {J. Phys.: Condens. Matter}\ }\textbf {\bibinfo {volume}
  {24}},\ \bibinfo {pages} {335401} (\bibinfo {year} {2012})}\BibitemShut
  {NoStop}%
\bibitem [{\citenamefont {Wang}\ and\ \citenamefont
  {Boercker}(1995)}]{WBJAP1995}%
  \BibitemOpen
  \bibfield  {author} {\bibinfo {author} {\bibfnamefont {Y.~R.}\ \bibnamefont
  {Wang}}\ and\ \bibinfo {author} {\bibfnamefont {D.~B.}\ \bibnamefont
  {Boercker}},\ }\bibfield  {title} {\enquote {\bibinfo {title} {Effective
  interatomic potential for body-centered-cubic metals},}\ }\href
  {https://doi.org/10.1063/1.360661} {\bibfield  {journal} {\bibinfo  {journal}
  {J. Appl. Phys.}\ }\textbf {\bibinfo {volume} {78}},\ \bibinfo {pages}
  {122--126} (\bibinfo {year} {1995})}\BibitemShut {NoStop}%
\bibitem [{\citenamefont {Johnson}\ and\ \citenamefont {Oh}(1989)}]{JOJMR1989}%
  \BibitemOpen
  \bibfield  {author} {\bibinfo {author} {\bibfnamefont {R.~A.}\ \bibnamefont
  {Johnson}}\ and\ \bibinfo {author} {\bibfnamefont {D.~J.}\ \bibnamefont
  {Oh}},\ }\bibfield  {title} {\enquote {\bibinfo {title} {Analytic embedded
  atom method model for bcc metals},}\ }\href
  {https://doi.org/10.1557/JMR.1989.1195} {\bibfield  {journal} {\bibinfo
  {journal} {Journal of Materials Research}\ }\textbf {\bibinfo {volume} {4}},\
  \bibinfo {pages} {1195--1201} (\bibinfo {year} {1989})}\BibitemShut {NoStop}%
\bibitem [{\citenamefont {Johnson}(1988)}]{JohnsonPRB1988}%
  \BibitemOpen
  \bibfield  {author} {\bibinfo {author} {\bibfnamefont {R.~A.}\ \bibnamefont
  {Johnson}},\ }\bibfield  {title} {\enquote {\bibinfo {title} {Analytic
  nearest-neighbor model for fcc metals},}\ }\href
  {https://doi.org/10.1103/PhysRevB.37.3924} {\bibfield  {journal} {\bibinfo
  {journal} {Phys. Rev. B}\ }\textbf {\bibinfo {volume} {37}},\ \bibinfo
  {pages} {3924--3931} (\bibinfo {year} {1988})}\BibitemShut {NoStop}%
\bibitem [{\citenamefont {Smith}\ \emph {et~al.}(1968)\citenamefont {Smith},
  \citenamefont {Dolling}, \citenamefont {Nicklow}, \citenamefont
  {Vijayaraghavan},\ and\ \citenamefont {Wilkinson}}]{SmithNIS1968}%
  \BibitemOpen
  \bibfield  {author} {\bibinfo {author} {\bibfnamefont {H.~G.}\ \bibnamefont
  {Smith}}, \bibinfo {author} {\bibfnamefont {G.}~\bibnamefont {Dolling}},
  \bibinfo {author} {\bibfnamefont {R.~M.}\ \bibnamefont {Nicklow}}, \bibinfo
  {author} {\bibfnamefont {P.~R.}\ \bibnamefont {Vijayaraghavan}}, \ and\
  \bibinfo {author} {\bibfnamefont {M.~K.}\ \bibnamefont {Wilkinson}},\
  }\bibfield  {title} {\enquote {\bibinfo {title} {Phonon dispersion curves in
  lithium},}\ }\href
  {https://inis.iaea.org/search/search.aspx?orig_q=RN:44068884} {\bibfield
  {journal} {\bibinfo  {journal} {Neutron Inelastic Scattering}\ }\textbf
  {\bibinfo {volume} {1}},\ \bibinfo {pages} {149--155} (\bibinfo {year}
  {1968})}\BibitemShut {NoStop}%
\bibitem [{\citenamefont {Woods}\ \emph {et~al.}(1962)\citenamefont {Woods},
  \citenamefont {Brockhouse}, \citenamefont {March}, \citenamefont {Stewart},\
  and\ \citenamefont {Bowers}}]{WoodsPR1962}%
  \BibitemOpen
  \bibfield  {author} {\bibinfo {author} {\bibfnamefont {A.~D.~B.}\
  \bibnamefont {Woods}}, \bibinfo {author} {\bibfnamefont {B.~N.}\ \bibnamefont
  {Brockhouse}}, \bibinfo {author} {\bibfnamefont {R.~H.}\ \bibnamefont
  {March}}, \bibinfo {author} {\bibfnamefont {A.~T.}\ \bibnamefont {Stewart}},
  \ and\ \bibinfo {author} {\bibfnamefont {R.}~\bibnamefont {Bowers}},\
  }\bibfield  {title} {\enquote {\bibinfo {title} {Crystal dynamics of sodium
  at 90$^\circ${K}},}\ }\href {https://doi.org/10.1103/PhysRev.128.1112}
  {\bibfield  {journal} {\bibinfo  {journal} {Phys. Rev.}\ }\textbf {\bibinfo
  {volume} {128}},\ \bibinfo {pages} {1112--1120} (\bibinfo {year}
  {1962})}\BibitemShut {NoStop}%
\bibitem [{\citenamefont {Cowley}\ \emph {et~al.}(1966)\citenamefont {Cowley},
  \citenamefont {Woods},\ and\ \citenamefont {Dolling}}]{CowleyPR1966}%
  \BibitemOpen
  \bibfield  {author} {\bibinfo {author} {\bibfnamefont {R.~A.}\ \bibnamefont
  {Cowley}}, \bibinfo {author} {\bibfnamefont {A.~D.~B.}\ \bibnamefont
  {Woods}}, \ and\ \bibinfo {author} {\bibfnamefont {G.}~\bibnamefont
  {Dolling}},\ }\bibfield  {title} {\enquote {\bibinfo {title} {Crystal
  dynamics of potassium. {I}. {P}seudopotential analysis of phonon dispersion
  curves at 9$^\circ${K}},}\ }\href {https://doi.org/10.1103/PhysRev.150.487}
  {\bibfield  {journal} {\bibinfo  {journal} {Phys. Rev.}\ }\textbf {\bibinfo
  {volume} {150}},\ \bibinfo {pages} {487} (\bibinfo {year}
  {1966})}\BibitemShut {NoStop}%
\bibitem [{\citenamefont {Copley}\ and\ \citenamefont
  {Brockhouse}(1973)}]{CopleyCJP1973}%
  \BibitemOpen
  \bibfield  {author} {\bibinfo {author} {\bibfnamefont {J.~R.~D.}\
  \bibnamefont {Copley}}\ and\ \bibinfo {author} {\bibfnamefont {B.~N.}\
  \bibnamefont {Brockhouse}},\ }\bibfield  {title} {\enquote {\bibinfo {title}
  {Crystal dynamics of rubidium. {I}. {M}easurements and harmonic analysis},}\
  }\href {https://doi.org/10.1139/p73-087} {\bibfield  {journal} {\bibinfo
  {journal} {Canadian Journal of Physics}\ }\textbf {\bibinfo {volume} {51}},\
  \bibinfo {pages} {657} (\bibinfo {year} {1973})}\BibitemShut {NoStop}%
\bibitem [{\citenamefont {Kittel}(2005)}]{Kittel2005}%
  \BibitemOpen
  \bibfield  {author} {\bibinfo {author} {\bibfnamefont {C.}~\bibnamefont
  {Kittel}},\ }\href@noop {} {\emph {\bibinfo {title} {Introduction to Solid
  State Physics}}}\ (\bibinfo  {publisher} {John Wiley \& Sons, New York},\
  \bibinfo {year} {2005})\BibitemShut {NoStop}%
\bibitem [{\citenamefont {MacDonald}(1953)}]{MacDonaldJCP1953}%
  \BibitemOpen
  \bibfield  {author} {\bibinfo {author} {\bibfnamefont {D.~K.~C.}\
  \bibnamefont {MacDonald}},\ }\bibfield  {title} {\enquote {\bibinfo {title}
  {Self-diffusion in the alkali metals},}\ }\href {\doibase 10.1063/1.1698594}
  {\bibfield  {journal} {\bibinfo  {journal} {J. Chem. Phys.}\ }\textbf
  {\bibinfo {volume} {21}},\ \bibinfo {pages} {177} (\bibinfo {year}
  {1953})}\BibitemShut {NoStop}%
\bibitem [{\citenamefont {Adlhart}\ \emph {et~al.}(1975)\citenamefont
  {Adlhart}, \citenamefont {Fritsch},\ and\ \citenamefont
  {L{\"u}scher}}]{adlhart1975}%
  \BibitemOpen
  \bibfield  {author} {\bibinfo {author} {\bibfnamefont {W}~\bibnamefont
  {Adlhart}}, \bibinfo {author} {\bibfnamefont {G}~\bibnamefont {Fritsch}}, \
  and\ \bibinfo {author} {\bibfnamefont {E}~\bibnamefont {L{\"u}scher}},\
  }\bibfield  {title} {\enquote {\bibinfo {title} {Equilibrium defect
  properties of sodium in the high temperature range},}\ }\href
  {https://doi.org/10.1016/0022-3697(75)90224-3} {\bibfield  {journal}
  {\bibinfo  {journal} {J. Phys. Chem. Solids}\ }\textbf {\bibinfo {volume}
  {36}},\ \bibinfo {pages} {1405--1409} (\bibinfo {year} {1975})}\BibitemShut
  {NoStop}%
\bibitem [{\citenamefont {Mundy}\ \emph {et~al.}(1971)\citenamefont {Mundy},
  \citenamefont {Miller},\ and\ \citenamefont {Porte}}]{MundyPRB1971}%
  \BibitemOpen
  \bibfield  {author} {\bibinfo {author} {\bibfnamefont {J.~N.}\ \bibnamefont
  {Mundy}}, \bibinfo {author} {\bibfnamefont {T.~E.}\ \bibnamefont {Miller}}, \
  and\ \bibinfo {author} {\bibfnamefont {R.~J.}\ \bibnamefont {Porte}},\
  }\bibfield  {title} {\enquote {\bibinfo {title} {Self-diffusion in
  potassium},}\ }\href {https://doi.org/10.1103/PhysRevB.3.2445} {\bibfield
  {journal} {\bibinfo  {journal} {Phys. Rev.}\ }\textbf {\bibinfo {volume}
  {3}},\ \bibinfo {pages} {2445--2447} (\bibinfo {year} {1971})}\BibitemShut
  {NoStop}%
\bibitem [{\citenamefont {Martin}(1965)}]{MartinPR1965}%
  \BibitemOpen
  \bibfield  {author} {\bibinfo {author} {\bibfnamefont {Douglas~L.}\
  \bibnamefont {Martin}},\ }\bibfield  {title} {\enquote {\bibinfo {title}
  {Analysis of alkali-metal specific-heat data},}\ }\href
  {https://doi.org/10.1103/PhysRev.139.A150} {\bibfield  {journal} {\bibinfo
  {journal} {Phys. Rev.}\ }\textbf {\bibinfo {volume} {139}},\ \bibinfo {pages}
  {A150} (\bibinfo {year} {1965})}\BibitemShut {NoStop}%
\bibitem [{\citenamefont {Anderson}\ and\ \citenamefont
  {Swenson}(1985)}]{AndersonPRB1985}%
  \BibitemOpen
  \bibfield  {author} {\bibinfo {author} {\bibfnamefont {M.~S.}\ \bibnamefont
  {Anderson}}\ and\ \bibinfo {author} {\bibfnamefont {C.~A.}\ \bibnamefont
  {Swenson}},\ }\bibfield  {title} {\enquote {\bibinfo {title} {Experimental
  equations of state for cesium and lithium metals to 20 kbar and the
  high-pressure behavior of the alkali metals},}\ }\href
  {https://doi.org/10.1103/PhysRevB.31.668} {\bibfield  {journal} {\bibinfo
  {journal} {Phys. Rev. B}\ }\textbf {\bibinfo {volume} {31}},\ \bibinfo
  {pages} {668} (\bibinfo {year} {1985})}\BibitemShut {NoStop}%
\bibitem [{\citenamefont {Siegel}\ and\ \citenamefont
  {Quimby}(1938)}]{SiegelPR1938}%
  \BibitemOpen
  \bibfield  {author} {\bibinfo {author} {\bibfnamefont {S.}~\bibnamefont
  {Siegel}}\ and\ \bibinfo {author} {\bibfnamefont {S.~L.}\ \bibnamefont
  {Quimby}},\ }\bibfield  {title} {\enquote {\bibinfo {title} {The thermal
  expansion of crystalline sodium between 80$^\circ${K} and 290$^\circ${K}},}\
  }\href {https://doi.org/10.1103/PhysRev.54.76} {\bibfield  {journal}
  {\bibinfo  {journal} {Phys. Rev.}\ }\textbf {\bibinfo {volume} {54}},\
  \bibinfo {pages} {76--78} (\bibinfo {year} {1938})}\BibitemShut {NoStop}%
\bibitem [{\citenamefont {Schouten}\ and\ \citenamefont
  {Swenson}(1974)}]{SchoutenPRB1974}%
  \BibitemOpen
  \bibfield  {author} {\bibinfo {author} {\bibfnamefont {D.~R.}\ \bibnamefont
  {Schouten}}\ and\ \bibinfo {author} {\bibfnamefont {C.~A.}\ \bibnamefont
  {Swenson}},\ }\bibfield  {title} {\enquote {\bibinfo {title}
  {Linear-thermal-expansion measurements on potassium metal from 2 to 320
  {K}},}\ }\href {https://doi.org/10.1103/PhysRevB.10.2175} {\bibfield
  {journal} {\bibinfo  {journal} {Phys. Rev. B}\ }\textbf {\bibinfo {volume}
  {10}},\ \bibinfo {pages} {2175--2185} (\bibinfo {year} {1974})}\BibitemShut
  {NoStop}%
\bibitem [{\citenamefont {Slotwinski}\ and\ \citenamefont
  {Trivisonno}(1969)}]{SlotwinskiJPCS1969}%
  \BibitemOpen
  \bibfield  {author} {\bibinfo {author} {\bibfnamefont {T.}~\bibnamefont
  {Slotwinski}}\ and\ \bibinfo {author} {\bibfnamefont {J.}~\bibnamefont
  {Trivisonno}},\ }\bibfield  {title} {\enquote {\bibinfo {title} {Temperature
  dependence of the elastic constants of single crystal lithium},}\ }\href
  {https://doi.org/10.1016/0022-3697(69)90386-2} {\bibfield  {journal}
  {\bibinfo  {journal} {J. Phys. Chem. Solids}\ }\textbf {\bibinfo {volume}
  {30}},\ \bibinfo {pages} {1276--1278} (\bibinfo {year} {1969})}\BibitemShut
  {NoStop}%
\bibitem [{\citenamefont {Quimby}\ and\ \citenamefont
  {Siegel}(1938)}]{QuimbyPR1938}%
  \BibitemOpen
  \bibfield  {author} {\bibinfo {author} {\bibfnamefont {S.~L.}\ \bibnamefont
  {Quimby}}\ and\ \bibinfo {author} {\bibfnamefont {S.}~\bibnamefont
  {Siegel}},\ }\bibfield  {title} {\enquote {\bibinfo {title} {The variation of
  the elastic constants of crystalline sodium with temperature between
  80$^\circ${K} and 210$^\circ${K}},}\ }\href
  {https://doi.org/10.1103/PhysRev.54.293} {\bibfield  {journal} {\bibinfo
  {journal} {Phys. Rev.}\ }\textbf {\bibinfo {volume} {54}},\ \bibinfo {pages}
  {293--299} (\bibinfo {year} {1938})}\BibitemShut {NoStop}%
\bibitem [{\citenamefont {Martinson}(1969)}]{MartinsonPR1969}%
  \BibitemOpen
  \bibfield  {author} {\bibinfo {author} {\bibfnamefont {R.~H.}\ \bibnamefont
  {Martinson}},\ }\bibfield  {title} {\enquote {\bibinfo {title} {Variation of
  the elastic constants of sodium with temperature and pressure},}\ }\href
  {https://doi.org/10.1103/PhysRev.178.902} {\bibfield  {journal} {\bibinfo
  {journal} {Phys. Rev.}\ }\textbf {\bibinfo {volume} {178}},\ \bibinfo {pages}
  {902--913} (\bibinfo {year} {1969})}\BibitemShut {NoStop}%
\bibitem [{\citenamefont {Marquardt}\ and\ \citenamefont
  {Trivisonno}(1965)}]{MarquardtJPCS1965}%
  \BibitemOpen
  \bibfield  {author} {\bibinfo {author} {\bibfnamefont {W.~R.}\ \bibnamefont
  {Marquardt}}\ and\ \bibinfo {author} {\bibfnamefont {J.}~\bibnamefont
  {Trivisonno}},\ }\bibfield  {title} {\enquote {\bibinfo {title} {Low
  temperature elastic constants of potassium},}\ }\href
  {https://doi.org/10.1016/0022-3697(65)90155-1} {\bibfield  {journal}
  {\bibinfo  {journal} {J. Phys. Chem. Solids}\ }\textbf {\bibinfo {volume}
  {26}},\ \bibinfo {pages} {273--278} (\bibinfo {year} {1965})}\BibitemShut
  {NoStop}%
\bibitem [{\citenamefont {Gutman}\ and\ \citenamefont
  {Trivisonno}(1967)}]{GutmanJPCS1967}%
  \BibitemOpen
  \bibfield  {author} {\bibinfo {author} {\bibfnamefont {E.~J.}\ \bibnamefont
  {Gutman}}\ and\ \bibinfo {author} {\bibfnamefont {J.}~\bibnamefont
  {Trivisonno}},\ }\bibfield  {title} {\enquote {\bibinfo {title} {Temperature
  dependence of the elastic constants of rubidium},}\ }\href
  {https://doi.org/10.1016/0022-3697(67)90009-1} {\bibfield  {journal}
  {\bibinfo  {journal} {J. Phys. Chem. Solids}\ }\textbf {\bibinfo {volume}
  {28}},\ \bibinfo {pages} {805--809} (\bibinfo {year} {1967})}\BibitemShut
  {NoStop}%
\bibitem [{\citenamefont {Smith}\ \emph {et~al.}(1991)\citenamefont {Smith},
  \citenamefont {Berliner}, \citenamefont {Jorgensen},\ and\ \citenamefont
  {Trivisonno}}]{smith1991}%
  \BibitemOpen
  \bibfield  {author} {\bibinfo {author} {\bibfnamefont {H.~G.}\ \bibnamefont
  {Smith}}, \bibinfo {author} {\bibfnamefont {R.}~\bibnamefont {Berliner}},
  \bibinfo {author} {\bibfnamefont {J.~D.}\ \bibnamefont {Jorgensen}}, \ and\
  \bibinfo {author} {\bibfnamefont {J.}~\bibnamefont {Trivisonno}},\ }\bibfield
   {title} {\enquote {\bibinfo {title} {Pressure effects on the martensitic
  transformation in metallic sodium},}\ }\href
  {https://doi.org/10.1103/PhysRevB.43.4524} {\bibfield  {journal} {\bibinfo
  {journal} {Phys. Rev. B}\ }\textbf {\bibinfo {volume} {43}},\ \bibinfo
  {pages} {4524--4526} (\bibinfo {year} {1991})}\BibitemShut {NoStop}%
\bibitem [{\citenamefont {Wang}\ \emph {et~al.}(2004)\citenamefont {Wang},
  \citenamefont {Curtarolo}, \citenamefont {Jiang}, \citenamefont {Arroyave},
  \citenamefont {Wang}, \citenamefont {Ceder}, \citenamefont {Chen},\ and\
  \citenamefont {Liu}}]{wang2004}%
  \BibitemOpen
  \bibfield  {author} {\bibinfo {author} {\bibfnamefont {Y.}~\bibnamefont
  {Wang}}, \bibinfo {author} {\bibfnamefont {S.}~\bibnamefont {Curtarolo}},
  \bibinfo {author} {\bibfnamefont {C.}~\bibnamefont {Jiang}}, \bibinfo
  {author} {\bibfnamefont {R.}~\bibnamefont {Arroyave}}, \bibinfo {author}
  {\bibfnamefont {T.}~\bibnamefont {Wang}}, \bibinfo {author} {\bibfnamefont
  {G.}~\bibnamefont {Ceder}}, \bibinfo {author} {\bibfnamefont {L.-Q.}\
  \bibnamefont {Chen}}, \ and\ \bibinfo {author} {\bibfnamefont {Z.-K.}\
  \bibnamefont {Liu}},\ }\bibfield  {title} {\enquote {\bibinfo {title} {Ab
  initio lattice stability in comparison with {CALPHAD} lattice stability},}\
  }\href {https://doi.org/10.1016/j.calphad.2004.05.002} {\bibfield  {journal}
  {\bibinfo  {journal} {Calphad}\ }\textbf {\bibinfo {volume} {28}},\ \bibinfo
  {pages} {79--90} (\bibinfo {year} {2004})}\BibitemShut {NoStop}%
\bibitem [{\citenamefont {Kulkarni}\ \emph {et~al.}(2011)\citenamefont
  {Kulkarni}, \citenamefont {Doll}, \citenamefont {Prasad}, \citenamefont
  {Sch{\"o}n},\ and\ \citenamefont {Jansen}}]{kulkarni2011}%
  \BibitemOpen
  \bibfield  {author} {\bibinfo {author} {\bibfnamefont {A.}~\bibnamefont
  {Kulkarni}}, \bibinfo {author} {\bibfnamefont {K.}~\bibnamefont {Doll}},
  \bibinfo {author} {\bibfnamefont {D.~L. V.~K.}\ \bibnamefont {Prasad}},
  \bibinfo {author} {\bibfnamefont {J.~C.}\ \bibnamefont {Sch{\"o}n}}, \ and\
  \bibinfo {author} {\bibfnamefont {M.}~\bibnamefont {Jansen}},\ }\bibfield
  {title} {\enquote {\bibinfo {title} {Alternative structure predicted for
  lithium at ambient pressure},}\ }\href
  {https://doi.org/10.1103/PhysRevB.84.172101} {\bibfield  {journal} {\bibinfo
  {journal} {Phys. Rev. B}\ }\textbf {\bibinfo {volume} {84}},\ \bibinfo
  {pages} {172101} (\bibinfo {year} {2011})}\BibitemShut {NoStop}%
\bibitem [{\citenamefont {Gaissmaier}\ \emph {et~al.}(2020)\citenamefont
  {Gaissmaier}, \citenamefont {van~den Borg}, \citenamefont {Fantauzzi},\ and\
  \citenamefont {Jacob}}]{gaissmaier2020}%
  \BibitemOpen
  \bibfield  {author} {\bibinfo {author} {\bibfnamefont {D.}~\bibnamefont
  {Gaissmaier}}, \bibinfo {author} {\bibfnamefont {M.}~\bibnamefont {van~den
  Borg}}, \bibinfo {author} {\bibfnamefont {D.}~\bibnamefont {Fantauzzi}}, \
  and\ \bibinfo {author} {\bibfnamefont {T.}~\bibnamefont {Jacob}},\ }\bibfield
   {title} {\enquote {\bibinfo {title} {Microscopic properties of {Na} and
  {Li}---a first principle study of metal battery anode materials},}\ }\href
  {https://doi.org/10.1002/cssc.201902860} {\bibfield  {journal} {\bibinfo
  {journal} {ChemSusChem}\ }\textbf {\bibinfo {volume} {13}},\ \bibinfo {pages}
  {771--783} (\bibinfo {year} {2020})}\BibitemShut {NoStop}%
\bibitem [{\citenamefont {Guellil}\ and\ \citenamefont
  {Adams}(1992)}]{GAJMR1992}%
  \BibitemOpen
  \bibfield  {author} {\bibinfo {author} {\bibfnamefont {A.~M.}\ \bibnamefont
  {Guellil}}\ and\ \bibinfo {author} {\bibfnamefont {J.~B.}\ \bibnamefont
  {Adams}},\ }\href {https://doi.org/10.1557/JMR.1992.0639} {\bibfield
  {journal} {\bibinfo  {journal} {Journal of Materials Research}\ }\textbf
  {\bibinfo {volume} {7}},\ \bibinfo {pages} {639--652} (\bibinfo {year}
  {1992})}\BibitemShut {NoStop}%
\bibitem [{\citenamefont {Chantasiriwan}\ and\ \citenamefont
  {Milstein}(1998)}]{CMPRB1998}%
  \BibitemOpen
  \bibfield  {author} {\bibinfo {author} {\bibfnamefont {S.}~\bibnamefont
  {Chantasiriwan}}\ and\ \bibinfo {author} {\bibfnamefont {F.}~\bibnamefont
  {Milstein}},\ }\bibfield  {title} {\enquote {\bibinfo {title} {Embedded-atom
  models of 12 cubic metals incorporating second- and third-order
  elastic-moduli data},}\ }\href {https://doi.org/10.1103/PhysRevB.58.5996}
  {\bibfield  {journal} {\bibinfo  {journal} {Phys. Rev. B}\ }\textbf {\bibinfo
  {volume} {58}},\ \bibinfo {pages} {5996--6005} (\bibinfo {year}
  {1998})}\BibitemShut {NoStop}%
\bibitem [{\citenamefont {H.}\ and\ \citenamefont {M.}(2002)}]{HuMSMSE2002}%
  \BibitemOpen
  \bibfield  {author} {\bibinfo {author} {\bibfnamefont {Wangyu}\ \bibnamefont
  {H.}}\ and\ \bibinfo {author} {\bibfnamefont {Fukumoto}\ \bibnamefont {M.}},\
  }\bibfield  {title} {\enquote {\bibinfo {title} {The application of the
  analytic embedded atom potentials to alkali metals},}\ }\href
  {https://doi.org/10.1088/0965-0393/10/6/307} {\bibfield  {journal} {\bibinfo
  {journal} {Modelling Simul. Mater. Sci. Eng.}\ }\textbf {\bibinfo {volume}
  {10}},\ \bibinfo {pages} {707--726} (\bibinfo {year} {2002})}\BibitemShut
  {NoStop}%
\bibitem [{\citenamefont {Xie}\ and\ \citenamefont {Zhang}(2008)}]{XieCJP2008}%
  \BibitemOpen
  \bibfield  {author} {\bibinfo {author} {\bibfnamefont {Y.}~\bibnamefont
  {Xie}}\ and\ \bibinfo {author} {\bibfnamefont {J.-M.}\ \bibnamefont
  {Zhang}},\ }\bibfield  {title} {\enquote {\bibinfo {title} {Atomistic
  simulation of phonon dispersion for body-centered cubic alkali metals},}\
  }\href {https://doi.org/10.1139/p07-200} {\bibfield  {journal} {\bibinfo
  {journal} {Canadian Journal of Physics}\ }\textbf {\bibinfo {volume} {86}},\
  \bibinfo {pages} {801--805} (\bibinfo {year} {2008})}\BibitemShut {NoStop}%
\bibitem [{\citenamefont {Zhang}\ \emph {et~al.}(2008)\citenamefont {Zhang},
  \citenamefont {Zhang},\ and\ \citenamefont {Xu}}]{ZhangJLTP2008}%
  \BibitemOpen
  \bibfield  {author} {\bibinfo {author} {\bibfnamefont {Jian-Min}\
  \bibnamefont {Zhang}}, \bibinfo {author} {\bibfnamefont {Xiao-Jun}\
  \bibnamefont {Zhang}}, \ and\ \bibinfo {author} {\bibfnamefont {Ke-Wei}\
  \bibnamefont {Xu}},\ }\bibfield  {title} {\enquote {\bibinfo {title} {{MAEAM}
  investigation of phonons for alkali metals},}\ }\href
  {https://doi.org/10.1007/s10909-007-9606-4} {\bibfield  {journal} {\bibinfo
  {journal} {J. Low Temp. Phys.}\ }\textbf {\bibinfo {volume} {150}},\ \bibinfo
  {pages} {730--738} (\bibinfo {year} {2008})}\BibitemShut {NoStop}%
\bibitem [{\citenamefont {Nichol}\ and\ \citenamefont
  {Ackland}(2016)}]{nichol2016}%
  \BibitemOpen
  \bibfield  {author} {\bibinfo {author} {\bibfnamefont {A.}~\bibnamefont
  {Nichol}}\ and\ \bibinfo {author} {\bibfnamefont {G.~J.}\ \bibnamefont
  {Ackland}},\ }\bibfield  {title} {\enquote {\bibinfo {title} {Property trends
  in simple metals: {A}n empirical potential approach},}\ }\href
  {https://doi.org/10.1103/PhysRevB.93.184101} {\bibfield  {journal} {\bibinfo
  {journal} {Phys. Rev. B}\ }\textbf {\bibinfo {volume} {93}},\ \bibinfo
  {pages} {184101} (\bibinfo {year} {2016})}\BibitemShut {NoStop}%
\bibitem [{\citenamefont {Ko}\ and\ \citenamefont {Jeon}(2017)}]{ko2017}%
  \BibitemOpen
  \bibfield  {author} {\bibinfo {author} {\bibfnamefont {W.-S.}\ \bibnamefont
  {Ko}}\ and\ \bibinfo {author} {\bibfnamefont {J.~B.}\ \bibnamefont {Jeon}},\
  }\bibfield  {title} {\enquote {\bibinfo {title} {Interatomic potential that
  describes martensitic phase transformations in pure lithium},}\ }\href
  {https://doi.org/10.1016/j.commatsci.2016.12.018} {\bibfield  {journal}
  {\bibinfo  {journal} {Computational Materials Science}\ }\textbf {\bibinfo
  {volume} {129}},\ \bibinfo {pages} {202--210} (\bibinfo {year}
  {2017})}\BibitemShut {NoStop}%
\bibitem [{\citenamefont {Kim}\ \emph {et~al.}(2020)\citenamefont {Kim},
  \citenamefont {Ko},\ and\ \citenamefont {Lee}}]{kim2020}%
  \BibitemOpen
  \bibfield  {author} {\bibinfo {author} {\bibfnamefont {Y.}~\bibnamefont
  {Kim}}, \bibinfo {author} {\bibfnamefont {W.-S.}\ \bibnamefont {Ko}}, \ and\
  \bibinfo {author} {\bibfnamefont {B.-J.}\ \bibnamefont {Lee}},\ }\bibfield
  {title} {\enquote {\bibinfo {title} {Second nearest-neighbor modified
  embedded atom method interatomic potentials for the {Na} unary and {Na-Sn}
  binary systems},}\ }\href {https://doi.org/10.1016/j.commatsci.2020.109953}
  {\bibfield  {journal} {\bibinfo  {journal} {Computational Materials Science}\
  }\textbf {\bibinfo {volume} {185}},\ \bibinfo {pages} {109953} (\bibinfo
  {year} {2020})}\BibitemShut {NoStop}%
\bibitem [{\citenamefont {Qin}\ \emph {et~al.}(2022)\citenamefont {Qin},
  \citenamefont {Wang}, \citenamefont {Li}, \citenamefont {Wen}, \citenamefont
  {Yin},\ and\ \citenamefont {Wu}}]{qin2022}%
  \BibitemOpen
  \bibfield  {author} {\bibinfo {author} {\bibfnamefont {Z.}~\bibnamefont
  {Qin}}, \bibinfo {author} {\bibfnamefont {R.}~\bibnamefont {Wang}}, \bibinfo
  {author} {\bibfnamefont {S.}~\bibnamefont {Li}}, \bibinfo {author}
  {\bibfnamefont {T.}~\bibnamefont {Wen}}, \bibinfo {author} {\bibfnamefont
  {B.}~\bibnamefont {Yin}}, \ and\ \bibinfo {author} {\bibfnamefont
  {Z.}~\bibnamefont {Wu}},\ }\bibfield  {title} {\enquote {\bibinfo {title}
  {Meam interatomic potential for thermodynamic and mechanical properties of
  lithium allotropes},}\ }\href
  {https://doi.org/10.1016/j.commatsci.2022.111706} {\bibfield  {journal}
  {\bibinfo  {journal} {Computational Materials Science}\ }\textbf {\bibinfo
  {volume} {214}},\ \bibinfo {pages} {111706} (\bibinfo {year}
  {2022})}\BibitemShut {NoStop}%
\bibitem [{\citenamefont {Riffe}\ \emph {et~al.}(2018)\citenamefont {Riffe},
  \citenamefont {Christensen},\ and\ \citenamefont {Wilson}}]{riffe2018}%
  \BibitemOpen
  \bibfield  {author} {\bibinfo {author} {\bibfnamefont {D.~M.}\ \bibnamefont
  {Riffe}}, \bibinfo {author} {\bibfnamefont {J.~D.}\ \bibnamefont
  {Christensen}}, \ and\ \bibinfo {author} {\bibfnamefont {R.~B.}\ \bibnamefont
  {Wilson}},\ }\bibfield  {title} {\enquote {\bibinfo {title} {Vibrational
  dynamics within the embedded-atom-method formalism and the relationship to
  {Born--von-K{\'a}rm{\'a}n} force constants},}\ }\href
  {https://doi.org/10.1088/1361-648X/aae09f} {\bibfield  {journal} {\bibinfo
  {journal} {J. Phys.: Condens. Matter}\ }\textbf {\bibinfo {volume} {30}},\
  \bibinfo {pages} {455702} (\bibinfo {year} {2018})}\BibitemShut {NoStop}%
\bibitem [{\citenamefont {Yorikawa}\ and\ \citenamefont
  {Muramatsu}(2008)}]{yorikawa2008}%
  \BibitemOpen
  \bibfield  {author} {\bibinfo {author} {\bibfnamefont {H.}~\bibnamefont
  {Yorikawa}}\ and\ \bibinfo {author} {\bibfnamefont {S.}~\bibnamefont
  {Muramatsu}},\ }\bibfield  {title} {\enquote {\bibinfo {title} {Theoretical
  study of crystal and electronic structures of {BiI}$_3$},}\ }\href
  {https://doi.org/10.1088/0953-8984/20/32/325220} {\bibfield  {journal}
  {\bibinfo  {journal} {J. Phys.: Condens. Matter}\ }\textbf {\bibinfo {volume}
  {20}},\ \bibinfo {pages} {325220} (\bibinfo {year} {2008})}\BibitemShut
  {NoStop}%
\bibitem [{\citenamefont {Grimvall}(1981)}]{Grimvall1981}%
  \BibitemOpen
  \bibfield  {author} {\bibinfo {author} {\bibfnamefont {G.}~\bibnamefont
  {Grimvall}},\ }\href@noop {} {\emph {\bibinfo {title} {The Electron-Phonon
  Interaction in Metals}}}\ (\bibinfo  {publisher} {North Holland, New York},\
  \bibinfo {year} {1981})\BibitemShut {NoStop}%
\bibitem [{\citenamefont {Riffe}\ and\ \citenamefont
  {Wilson}(2023)}]{Riffe2023}%
  \BibitemOpen
  \bibfield  {author} {\bibinfo {author} {\bibfnamefont {D.~M.}\ \bibnamefont
  {Riffe}}\ and\ \bibinfo {author} {\bibfnamefont {R.~B.}\ \bibnamefont
  {Wilson}},\ }\bibfield  {title} {\enquote {\bibinfo {title} {Excitation and
  relaxation of nonthermal electron energy distributions in metals with
  application to gold},}\ }\href {https://doi.org/10.1103/PhysRevB.107.214309}
  {\bibfield  {journal} {\bibinfo  {journal} {Phys. Rev. B}\ }\textbf {\bibinfo
  {volume} {107}},\ \bibinfo {pages} {214309} (\bibinfo {year}
  {2023})}\BibitemShut {NoStop}%
\bibitem [{\citenamefont {Faglioni}\ \emph {et~al.}(2016)\citenamefont
  {Faglioni}, \citenamefont {Merinov},\ and\ \citenamefont
  {Goddard~III}}]{faglioni2016}%
  \BibitemOpen
  \bibfield  {author} {\bibinfo {author} {\bibfnamefont {F.}~\bibnamefont
  {Faglioni}}, \bibinfo {author} {\bibfnamefont {B.~V.}\ \bibnamefont
  {Merinov}}, \ and\ \bibinfo {author} {\bibfnamefont {W.~A.}\ \bibnamefont
  {Goddard~III}},\ }\bibfield  {title} {\enquote {\bibinfo {title}
  {Room-temperature lithium phases from density functional theory},}\ }\href
  {https://doi.org/10.1021/acs.jpcc.6b08168} {\bibfield  {journal} {\bibinfo
  {journal} {The Journal of Physical Chemistry C}\ }\textbf {\bibinfo {volume}
  {120}},\ \bibinfo {pages} {27104--27108} (\bibinfo {year}
  {2016})}\BibitemShut {NoStop}%
\bibitem [{\citenamefont {McQuarrie}(1976)}]{McQuarrie1976}%
  \BibitemOpen
  \bibfield  {author} {\bibinfo {author} {\bibfnamefont {D.~A.}\ \bibnamefont
  {McQuarrie}},\ }\href@noop {} {\emph {\bibinfo {title} {Statistical
  Mechanics}}}\ (\bibinfo  {publisher} {Harper Collins, New York},\ \bibinfo
  {year} {1976})\BibitemShut {NoStop}%
\bibitem [{\citenamefont {Krystian}\ and\ \citenamefont
  {Pichl}(2000)}]{krystian2000}%
  \BibitemOpen
  \bibfield  {author} {\bibinfo {author} {\bibfnamefont {M.}~\bibnamefont
  {Krystian}}\ and\ \bibinfo {author} {\bibfnamefont {W.}~\bibnamefont
  {Pichl}},\ }\bibfield  {title} {\enquote {\bibinfo {title} {In situ optical
  microscopy of the martensitic phase transformation of lithium},}\ }\href
  {https://doi.org/10.1103/PhysRevB.62.13956} {\bibfield  {journal} {\bibinfo
  {journal} {Phys. Rev. B}\ }\textbf {\bibinfo {volume} {62}},\ \bibinfo
  {pages} {13956--13962} (\bibinfo {year} {2000})}\BibitemShut {NoStop}%
\bibitem [{\citenamefont {Pichl}\ \emph {et~al.}(2003)\citenamefont {Pichl},
  \citenamefont {Krystian}, \citenamefont {Prem},\ and\ \citenamefont
  {Krexner}}]{pichl2003}%
  \BibitemOpen
  \bibfield  {author} {\bibinfo {author} {\bibfnamefont {W.}~\bibnamefont
  {Pichl}}, \bibinfo {author} {\bibfnamefont {M.}~\bibnamefont {Krystian}},
  \bibinfo {author} {\bibfnamefont {M.}~\bibnamefont {Prem}}, \ and\ \bibinfo
  {author} {\bibfnamefont {G.}~\bibnamefont {Krexner}},\ }\bibfield  {title}
  {\enquote {\bibinfo {title} {The martensite phase of high-purity lithium},}\
  }\href {https://doi.org/10.1051/jp4:20031073} {\bibfield  {journal} {\bibinfo
   {journal} {Journal de Physique IV France}\ }\textbf {\bibinfo {volume}
  {112}},\ \bibinfo {pages} {1095--1098} (\bibinfo {year} {2003})}\BibitemShut
  {NoStop}%
\bibitem [{\citenamefont {Gooding}\ and\ \citenamefont
  {Krumhansl}(1988)}]{gooding1988}%
  \BibitemOpen
  \bibfield  {author} {\bibinfo {author} {\bibfnamefont {R.~J.}\ \bibnamefont
  {Gooding}}\ and\ \bibinfo {author} {\bibfnamefont {J.~A.}\ \bibnamefont
  {Krumhansl}},\ }\bibfield  {title} {\enquote {\bibinfo {title} {Theory of the
  bcc-to-{9R} structural phase transformation of {Li}},}\ }\href
  {https://doi.org/10.1103/PhysRevB.38.1695} {\bibfield  {journal} {\bibinfo
  {journal} {Phys. Rev. B}\ }\textbf {\bibinfo {volume} {38}},\ \bibinfo
  {pages} {1695--1704} (\bibinfo {year} {1988})}\BibitemShut {NoStop}%
\bibitem [{\citenamefont {Blaschko}\ \emph {et~al.}(1999)\citenamefont
  {Blaschko}, \citenamefont {Dmitriev}, \citenamefont {Krexner},\ and\
  \citenamefont {Tol{\'e}dano}}]{blaschko1999}%
  \BibitemOpen
  \bibfield  {author} {\bibinfo {author} {\bibfnamefont {O.}~\bibnamefont
  {Blaschko}}, \bibinfo {author} {\bibfnamefont {V.}~\bibnamefont {Dmitriev}},
  \bibinfo {author} {\bibfnamefont {G.}~\bibnamefont {Krexner}}, \ and\
  \bibinfo {author} {\bibfnamefont {P.}~\bibnamefont {Tol{\'e}dano}},\
  }\bibfield  {title} {\enquote {\bibinfo {title} {Theory of the martensitic
  phase transformations in lithium and sodium},}\ }\href
  {https://doi.org/10.1103/PhysRevB.59.9095} {\bibfield  {journal} {\bibinfo
  {journal} {Phys. Rev. B}\ }\textbf {\bibinfo {volume} {59}},\ \bibinfo
  {pages} {9095--9112} (\bibinfo {year} {1999})}\BibitemShut {NoStop}%
\bibitem [{\citenamefont {Smith}\ \emph {et~al.}(1990)\citenamefont {Smith},
  \citenamefont {Berliner}, \citenamefont {Jorgensen}, \citenamefont
  {Nielsen},\ and\ \citenamefont {Trivisonno}}]{smith1990}%
  \BibitemOpen
  \bibfield  {author} {\bibinfo {author} {\bibfnamefont {H.~G.}\ \bibnamefont
  {Smith}}, \bibinfo {author} {\bibfnamefont {R.}~\bibnamefont {Berliner}},
  \bibinfo {author} {\bibfnamefont {J.~D.}\ \bibnamefont {Jorgensen}}, \bibinfo
  {author} {\bibfnamefont {M.}~\bibnamefont {Nielsen}}, \ and\ \bibinfo
  {author} {\bibfnamefont {J.}~\bibnamefont {Trivisonno}},\ }\bibfield  {title}
  {\enquote {\bibinfo {title} {Pressure effects on the martensitic
  transformation in metallic lithium},}\ }\href
  {https://doi.org/10.1103/PhysRevB.41.1231} {\bibfield  {journal} {\bibinfo
  {journal} {Phys. Rev. B}\ }\textbf {\bibinfo {volume} {41}},\ \bibinfo
  {pages} {1231--1234} (\bibinfo {year} {1990})}\BibitemShut {NoStop}%
\bibitem [{\citenamefont {Boettger}\ and\ \citenamefont
  {Albers}(1989)}]{boettger1989}%
  \BibitemOpen
  \bibfield  {author} {\bibinfo {author} {\bibfnamefont {J.~C.}\ \bibnamefont
  {Boettger}}\ and\ \bibinfo {author} {\bibfnamefont {R.~C.}\ \bibnamefont
  {Albers}},\ }\bibfield  {title} {\enquote {\bibinfo {title} {Structural phase
  stability in lithium to ultrahigh pressures},}\ }\href
  {https://doi.org/10.1103/PhysRevB.39.3010} {\bibfield  {journal} {\bibinfo
  {journal} {Phys. Rev. B}\ }\textbf {\bibinfo {volume} {39}},\ \bibinfo
  {pages} {3010--3014} (\bibinfo {year} {1989})}\BibitemShut {NoStop}%
\bibitem [{\citenamefont {Nobel}\ \emph {et~al.}(1992)\citenamefont {Nobel},
  \citenamefont {Trickey}, \citenamefont {Blaha},\ and\ \citenamefont
  {Schwarz}}]{nobel1992}%
  \BibitemOpen
  \bibfield  {author} {\bibinfo {author} {\bibfnamefont {J.~A.}\ \bibnamefont
  {Nobel}}, \bibinfo {author} {\bibfnamefont {S.~B.}\ \bibnamefont {Trickey}},
  \bibinfo {author} {\bibfnamefont {P.}~\bibnamefont {Blaha}}, \ and\ \bibinfo
  {author} {\bibfnamefont {K.}~\bibnamefont {Schwarz}},\ }\bibfield  {title}
  {\enquote {\bibinfo {title} {Low-pressure crystalline phases of lithium},}\
  }\href {\doibase https://doi.org/10.1103/PhysRevB.45.5012} {\bibfield
  {journal} {\bibinfo  {journal} {Phys. Rev. B}\ }\textbf {\bibinfo {volume}
  {45}},\ \bibinfo {pages} {5012--5014} (\bibinfo {year} {1992})}\BibitemShut
  {NoStop}%
\bibitem [{\citenamefont {Perdew}\ \emph {et~al.}(1992)\citenamefont {Perdew},
  \citenamefont {Chevary}, \citenamefont {Vosko}, \citenamefont {Jackson},
  \citenamefont {Pederson}, \citenamefont {Singh},\ and\ \citenamefont
  {Fiolhais}}]{perdew1992}%
  \BibitemOpen
  \bibfield  {author} {\bibinfo {author} {\bibfnamefont {J.~P.}\ \bibnamefont
  {Perdew}}, \bibinfo {author} {\bibfnamefont {J.~A.}\ \bibnamefont {Chevary}},
  \bibinfo {author} {\bibfnamefont {S.~H.}\ \bibnamefont {Vosko}}, \bibinfo
  {author} {\bibfnamefont {K.~A.}\ \bibnamefont {Jackson}}, \bibinfo {author}
  {\bibfnamefont {M.~R.}\ \bibnamefont {Pederson}}, \bibinfo {author}
  {\bibfnamefont {D.~J.}\ \bibnamefont {Singh}}, \ and\ \bibinfo {author}
  {\bibfnamefont {C.}~\bibnamefont {Fiolhais}},\ }\bibfield  {title} {\enquote
  {\bibinfo {title} {Atoms, molecules, solids, and surfaces: Applications of
  the generalized gradient approximation for exchange and correlation},}\
  }\href {https://doi.org/10.1103/PhysRevB.46.6671} {\bibfield  {journal}
  {\bibinfo  {journal} {Phys. Rev. B}\ }\textbf {\bibinfo {volume} {46}},\
  \bibinfo {pages} {6671--6687} (\bibinfo {year} {1992})}\BibitemShut {NoStop}%
\bibitem [{\citenamefont {Papaconstantopoulos}\ and\ \citenamefont
  {Singh}(1992)}]{papa1992}%
  \BibitemOpen
  \bibfield  {author} {\bibinfo {author} {\bibfnamefont {D.~A.}\ \bibnamefont
  {Papaconstantopoulos}}\ and\ \bibinfo {author} {\bibfnamefont {D.~J.}\
  \bibnamefont {Singh}},\ }\bibfield  {title} {\enquote {\bibinfo {title}
  {Reply to ``{C}omment on `{T}otal-energy calculations of solid {H, Li, Na, K,
  Rb, and Cs}' ''},}\ }\href {https://doi.org/10.1103/PhysRevB.45.7507}
  {\bibfield  {journal} {\bibinfo  {journal} {Phys. Rev. B}\ }\textbf {\bibinfo
  {volume} {45}},\ \bibinfo {pages} {7507--7508} (\bibinfo {year}
  {1992})}\BibitemShut {NoStop}%
\bibitem [{\citenamefont {Fiolhais}\ \emph {et~al.}(1995)\citenamefont
  {Fiolhais}, \citenamefont {Perdew}, \citenamefont {Armster}, \citenamefont
  {MacLaren},\ and\ \citenamefont {Brajczewska}}]{fiolhais1995}%
  \BibitemOpen
  \bibfield  {author} {\bibinfo {author} {\bibfnamefont {C.}~\bibnamefont
  {Fiolhais}}, \bibinfo {author} {\bibfnamefont {J.~P.}\ \bibnamefont
  {Perdew}}, \bibinfo {author} {\bibfnamefont {S.~Q.}\ \bibnamefont {Armster}},
  \bibinfo {author} {\bibfnamefont {J.~M.}\ \bibnamefont {MacLaren}}, \ and\
  \bibinfo {author} {\bibfnamefont {M.}~\bibnamefont {Brajczewska}},\
  }\bibfield  {title} {\enquote {\bibinfo {title} {Dominant density parameters
  and local pseudopotentials for simple metals},}\ }\href
  {https://doi.org/10.1103/PhysRevB.51.14001} {\bibfield  {journal} {\bibinfo
  {journal} {Phys. Rev. B}\ }\textbf {\bibinfo {volume} {51}},\ \bibinfo
  {pages} {14001--14011} (\bibinfo {year} {1995})}\BibitemShut {NoStop}%
\bibitem [{\citenamefont {Sliwko}\ \emph {et~al.}(1996)\citenamefont {Sliwko},
  \citenamefont {Mohn}, \citenamefont {Schwarz},\ and\ \citenamefont
  {Blaha}}]{sliwko1996}%
  \BibitemOpen
  \bibfield  {author} {\bibinfo {author} {\bibfnamefont {V.~L.}\ \bibnamefont
  {Sliwko}}, \bibinfo {author} {\bibfnamefont {P.}~\bibnamefont {Mohn}},
  \bibinfo {author} {\bibfnamefont {K.}~\bibnamefont {Schwarz}}, \ and\
  \bibinfo {author} {\bibfnamefont {P.}~\bibnamefont {Blaha}},\ }\bibfield
  {title} {\enquote {\bibinfo {title} {The fcc - bcc structural transition: I.
  {A} band theoretical study for {Li, K, Rb, Ca, Sr, and the transition metals
  Ti and V}},}\ }\href {https://doi.org/10.1088/0953-8984/8/7/006} {\bibfield
  {journal} {\bibinfo  {journal} {J. Phys.: Condens. Matter}\ }\textbf
  {\bibinfo {volume} {8}},\ \bibinfo {pages} {799--815} (\bibinfo {year}
  {1996})}\BibitemShut {NoStop}%
\bibitem [{\citenamefont {Doll}\ \emph {et~al.}(1999)\citenamefont {Doll},
  \citenamefont {Harrison},\ and\ \citenamefont {Saunders}}]{doll1999}%
  \BibitemOpen
  \bibfield  {author} {\bibinfo {author} {\bibfnamefont {K.}~\bibnamefont
  {Doll}}, \bibinfo {author} {\bibfnamefont {N.~M.}\ \bibnamefont {Harrison}},
  \ and\ \bibinfo {author} {\bibfnamefont {V~R.}\ \bibnamefont {Saunders}},\
  }\bibfield  {title} {\enquote {\bibinfo {title} {A density functional study
  of lithium bulk and surfaces},}\ }\href
  {https://doi.org/10.1088/0953-8984/11/26/305} {\bibfield  {journal} {\bibinfo
   {journal} {J. Phys.: Condens. Matter}\ }\textbf {\bibinfo {volume} {11}},\
  \bibinfo {pages} {5007} (\bibinfo {year} {1999})}\BibitemShut {NoStop}%
\bibitem [{\citenamefont {Hanfland}\ \emph {et~al.}(1999)\citenamefont
  {Hanfland}, \citenamefont {Loa}, \citenamefont {Syassen}, \citenamefont
  {Schwarz},\ and\ \citenamefont {Takemura}}]{hanfland1999}%
  \BibitemOpen
  \bibfield  {author} {\bibinfo {author} {\bibfnamefont {M.}~\bibnamefont
  {Hanfland}}, \bibinfo {author} {\bibfnamefont {I.}~\bibnamefont {Loa}},
  \bibinfo {author} {\bibfnamefont {K.}~\bibnamefont {Syassen}}, \bibinfo
  {author} {\bibfnamefont {U.}~\bibnamefont {Schwarz}}, \ and\ \bibinfo
  {author} {\bibfnamefont {K.}~\bibnamefont {Takemura}},\ }\bibfield  {title}
  {\enquote {\bibinfo {title} {Equation of state of lithium to 21 {GPa}},}\
  }\href {https://doi.org/10.1016/S0038-1098(99)00322-1} {\bibfield  {journal}
  {\bibinfo  {journal} {Solid State Commun.}\ }\textbf {\bibinfo {volume}
  {112}},\ \bibinfo {pages} {123--127} (\bibinfo {year} {1999})}\BibitemShut
  {NoStop}%
\bibitem [{\citenamefont {Xie}\ \emph {et~al.}(2008)\citenamefont {Xie},
  \citenamefont {Ma}, \citenamefont {Cui}, \citenamefont {Li}, \citenamefont
  {Qiu},\ and\ \citenamefont {Zou}}]{xie2008A}%
  \BibitemOpen
  \bibfield  {author} {\bibinfo {author} {\bibfnamefont {Y.}~\bibnamefont
  {Xie}}, \bibinfo {author} {\bibfnamefont {Y.~M.}\ \bibnamefont {Ma}},
  \bibinfo {author} {\bibfnamefont {T.}~\bibnamefont {Cui}}, \bibinfo {author}
  {\bibfnamefont {Y.}~\bibnamefont {Li}}, \bibinfo {author} {\bibfnamefont
  {J.}~\bibnamefont {Qiu}}, \ and\ \bibinfo {author} {\bibfnamefont {G.~T.}\
  \bibnamefont {Zou}},\ }\bibfield  {title} {\enquote {\bibinfo {title} {Origin
  of bcc to fcc phase transition under pressure in alkali metals},}\ }\href
  {https://doi.org/10.1088/1367-2630/10/6/063022} {\bibfield  {journal}
  {\bibinfo  {journal} {New Journal of Physics}\ }\textbf {\bibinfo {volume}
  {10}},\ \bibinfo {pages} {063022} (\bibinfo {year} {2008})}\BibitemShut
  {NoStop}%
\bibitem [{\citenamefont {Liang}\ and\ \citenamefont {Gong}(2010)}]{liang2010}%
  \BibitemOpen
  \bibfield  {author} {\bibinfo {author} {\bibfnamefont {C.~P.}\ \bibnamefont
  {Liang}}\ and\ \bibinfo {author} {\bibfnamefont {H.~R.}\ \bibnamefont
  {Gong}},\ }\bibfield  {title} {\enquote {\bibinfo {title} {Phase stability,
  mechanical property, and electronic structure of {Mg--Li} system},}\ }\href
  {https://doi.org/10.1016/j.jallcom.2009.09.032} {\bibfield  {journal}
  {\bibinfo  {journal} {Journal of Alloys and Compounds}\ }\textbf {\bibinfo
  {volume} {489}},\ \bibinfo {pages} {130--135} (\bibinfo {year}
  {2010})}\BibitemShut {NoStop}%
\bibitem [{\citenamefont {Cui}\ \emph {et~al.}(2012)\citenamefont {Cui},
  \citenamefont {Gao}, \citenamefont {Cui},\ and\ \citenamefont
  {Qu}}]{cui2012}%
  \BibitemOpen
  \bibfield  {author} {\bibinfo {author} {\bibfnamefont {Z.}~\bibnamefont
  {Cui}}, \bibinfo {author} {\bibfnamefont {F.}~\bibnamefont {Gao}}, \bibinfo
  {author} {\bibfnamefont {Z.}~\bibnamefont {Cui}}, \ and\ \bibinfo {author}
  {\bibfnamefont {J.}~\bibnamefont {Qu}},\ }\bibfield  {title} {\enquote
  {\bibinfo {title} {Developing a second nearest-neighbor modified embedded
  atom method interatomic potential for lithium},}\ }\href
  {https://doi.org/10.1088/0965-0393/20/1/015014} {\bibfield  {journal}
  {\bibinfo  {journal} {Modelling Simul. Mater. Sci. Eng.}\ }\textbf {\bibinfo
  {volume} {20}},\ \bibinfo {pages} {015014} (\bibinfo {year}
  {2012})}\BibitemShut {NoStop}%
\bibitem [{\citenamefont {Shin}\ and\ \citenamefont {Carter}(2014)}]{shin2014}%
  \BibitemOpen
  \bibfield  {author} {\bibinfo {author} {\bibfnamefont {I.}~\bibnamefont
  {Shin}}\ and\ \bibinfo {author} {\bibfnamefont {E.~A.}\ \bibnamefont
  {Carter}},\ }\bibfield  {title} {\enquote {\bibinfo {title} {First-principles
  simulations of plasticity in body-centered-cubic magnesium--lithium
  alloys},}\ }\href {https://doi.org/10.1016/j.actamat.2013.10.030} {\bibfield
  {journal} {\bibinfo  {journal} {Acta Materialia}\ }\textbf {\bibinfo {volume}
  {64}},\ \bibinfo {pages} {198--207} (\bibinfo {year} {2014})}\BibitemShut
  {NoStop}%
\bibitem [{\citenamefont {Legrain}\ and\ \citenamefont
  {Manzhos}(2015)}]{legrain2015}%
  \BibitemOpen
  \bibfield  {author} {\bibinfo {author} {\bibfnamefont {F.}~\bibnamefont
  {Legrain}}\ and\ \bibinfo {author} {\bibfnamefont {S.}~\bibnamefont
  {Manzhos}},\ }\bibfield  {title} {\enquote {\bibinfo {title} {Highly accurate
  local pseudopotentials of {Li, Na, and Mg} for orbital free density
  functional theory},}\ }\href {https://doi.org/10.1016/j.cplett.2015.01.016}
  {\bibfield  {journal} {\bibinfo  {journal} {Chem. Phys. Lett.}\ }\textbf
  {\bibinfo {volume} {622}},\ \bibinfo {pages} {99--103} (\bibinfo {year}
  {2015})}\BibitemShut {NoStop}%
\bibitem [{\citenamefont {Raju~Natarajan}\ and\ \citenamefont {Van~der
  Ven}(2019)}]{rajunatarajan2019}%
  \BibitemOpen
  \bibfield  {author} {\bibinfo {author} {\bibfnamefont {A.}~\bibnamefont
  {Raju~Natarajan}}\ and\ \bibinfo {author} {\bibfnamefont {A.}~\bibnamefont
  {Van~der Ven}},\ }\bibfield  {title} {\enquote {\bibinfo {title} {Toward an
  understanding of deformation mechanisms in metallic lithium and sodium from
  first-principles},}\ }\href {https://doi.org/10.1021/acs.chemmater.9b03422}
  {\bibfield  {journal} {\bibinfo  {journal} {Chemistry of Materials}\ }\textbf
  {\bibinfo {volume} {31}},\ \bibinfo {pages} {8222--8229} (\bibinfo {year}
  {2019})}\BibitemShut {NoStop}%
\bibitem [{\citenamefont {Moriarty}\ and\ \citenamefont
  {McMahan}(1982)}]{moriarty1982}%
  \BibitemOpen
  \bibfield  {author} {\bibinfo {author} {\bibfnamefont {J.~A.}\ \bibnamefont
  {Moriarty}}\ and\ \bibinfo {author} {\bibfnamefont {A.~K.}\ \bibnamefont
  {McMahan}},\ }\bibfield  {title} {\enquote {\bibinfo {title} {High-pressure
  structural phase transitions in {Na, Mg, and Al}},}\ }\href
  {https://doi.org/10.1103/PhysRevLett.48.809} {\bibfield  {journal} {\bibinfo
  {journal} {Phys. Rev. Lett.}\ }\textbf {\bibinfo {volume} {48}},\ \bibinfo
  {pages} {809--812} (\bibinfo {year} {1982})}\BibitemShut {NoStop}%
\bibitem [{\citenamefont {Ye}\ \emph {et~al.}(1990)\citenamefont {Ye},
  \citenamefont {Chan}, \citenamefont {Ho},\ and\ \citenamefont
  {Harmon}}]{ye1990}%
  \BibitemOpen
  \bibfield  {author} {\bibinfo {author} {\bibfnamefont {Y.-Y}\ \bibnamefont
  {Ye}}, \bibinfo {author} {\bibfnamefont {C.-T.}\ \bibnamefont {Chan}},
  \bibinfo {author} {\bibfnamefont {K.-M.}\ \bibnamefont {Ho}}, \ and\ \bibinfo
  {author} {\bibfnamefont {B.~N.}\ \bibnamefont {Harmon}},\ }\bibfield  {title}
  {\enquote {\bibinfo {title} {Total energy calculations for structural phase
  transformations},}\ }\href {https://doi.org/10.1177/109434209000400311}
  {\bibfield  {journal} {\bibinfo  {journal} {The International Journal of
  Supercomputing Applications}\ }\textbf {\bibinfo {volume} {4}},\ \bibinfo
  {pages} {111--121} (\bibinfo {year} {1990})}\BibitemShut {NoStop}%
\bibitem [{\citenamefont {Nogueira}\ \emph {et~al.}(1999)\citenamefont
  {Nogueira}, \citenamefont {Fiolhais},\ and\ \citenamefont
  {Perdew}}]{nogueira1999}%
  \BibitemOpen
  \bibfield  {author} {\bibinfo {author} {\bibfnamefont {F.}~\bibnamefont
  {Nogueira}}, \bibinfo {author} {\bibfnamefont {C.}~\bibnamefont {Fiolhais}},
  \ and\ \bibinfo {author} {\bibfnamefont {J.~P.}\ \bibnamefont {Perdew}},\
  }\bibfield  {title} {\enquote {\bibinfo {title} {Trends in the properties and
  structures of the simple metals from a universal local pseudopotential},}\
  }\href {https://doi.org/10.1103/PhysRevB.59.2570} {\bibfield  {journal}
  {\bibinfo  {journal} {Phys. Rev. B}\ }\textbf {\bibinfo {volume} {59}},\
  \bibinfo {pages} {2570--2578} (\bibinfo {year} {1999})}\BibitemShut {NoStop}%
\bibitem [{\citenamefont {Katsnelson}\ \emph {et~al.}(2000)\citenamefont
  {Katsnelson}, \citenamefont {Sinko}, \citenamefont {Smirnov}, \citenamefont
  {Trefilov},\ and\ \citenamefont {Khromov}}]{katsnelson2000}%
  \BibitemOpen
  \bibfield  {author} {\bibinfo {author} {\bibfnamefont {M.~I.}\ \bibnamefont
  {Katsnelson}}, \bibinfo {author} {\bibfnamefont {G.~V.}\ \bibnamefont
  {Sinko}}, \bibinfo {author} {\bibfnamefont {N.~A.}\ \bibnamefont {Smirnov}},
  \bibinfo {author} {\bibfnamefont {A.~V.}\ \bibnamefont {Trefilov}}, \ and\
  \bibinfo {author} {\bibfnamefont {K.~Yu.}\ \bibnamefont {Khromov}},\
  }\bibfield  {title} {\enquote {\bibinfo {title} {Structure, elastic moduli,
  and thermodynamics of sodium and potassium at ultrahigh pressures},}\ }\href
  {https://doi.org/10.1103/PhysRevB.61.14420} {\bibfield  {journal} {\bibinfo
  {journal} {Phys. Rev. B}\ }\textbf {\bibinfo {volume} {61}},\ \bibinfo
  {pages} {14420--14424} (\bibinfo {year} {2000})}\BibitemShut {NoStop}%
\bibitem [{\citenamefont {Bonsignori}\ and\ \citenamefont
  {Magnaterra}(1982)}]{bonsignori1982}%
  \BibitemOpen
  \bibfield  {author} {\bibinfo {author} {\bibfnamefont {F.}~\bibnamefont
  {Bonsignori}}\ and\ \bibinfo {author} {\bibfnamefont {A.}~\bibnamefont
  {Magnaterra}},\ }\bibfield  {title} {\enquote {\bibinfo {title} {The
  pseudopotential of heavy alkali metals: {N}onlocality and d-level effects on
  the structure stability},}\ }\href {https://doi.org/10.1007/BF02451069}
  {\bibfield  {journal} {\bibinfo  {journal} {Il Nuovo Cimento D}\ }\textbf
  {\bibinfo {volume} {1}},\ \bibinfo {pages} {789--801} (\bibinfo {year}
  {1982})}\BibitemShut {NoStop}%
\bibitem [{\citenamefont {Rahman}\ \emph {et~al.}(1984)\citenamefont {Rahman},
  \citenamefont {Mujibar~Rahman},\ and\ \citenamefont
  {Chowdhury}}]{rahman1984}%
  \BibitemOpen
  \bibfield  {author} {\bibinfo {author} {\bibfnamefont {M.~M.}\ \bibnamefont
  {Rahman}}, \bibinfo {author} {\bibfnamefont {S.~M.}\ \bibnamefont
  {Mujibar~Rahman}}, \ and\ \bibinfo {author} {\bibfnamefont {S.~M. M.~R.}\
  \bibnamefont {Chowdhury}},\ }\bibfield  {title} {\enquote {\bibinfo {title}
  {Structural energetics of heavy alkali metals: pseudopotential theory
  revisited},}\ }\href {https://doi.org/10.1088/0305-4608/14/4/008} {\bibfield
  {journal} {\bibinfo  {journal} {Journal of Physics F: Metal Physics}\
  }\textbf {\bibinfo {volume} {14}},\ \bibinfo {pages} {833--} (\bibinfo {year}
  {1984})}\BibitemShut {NoStop}%
\bibitem [{\citenamefont {Alouani}\ \emph {et~al.}(1989)\citenamefont
  {Alouani}, \citenamefont {Christensen},\ and\ \citenamefont
  {Syassen}}]{alouani1989}%
  \BibitemOpen
  \bibfield  {author} {\bibinfo {author} {\bibfnamefont {M.}~\bibnamefont
  {Alouani}}, \bibinfo {author} {\bibfnamefont {N.~E.}\ \bibnamefont
  {Christensen}}, \ and\ \bibinfo {author} {\bibfnamefont {K.}~\bibnamefont
  {Syassen}},\ }\bibfield  {title} {\enquote {\bibinfo {title} {Calculated
  ground-state and optical properties of potassium under pressure},}\ }\href
  {https://doi.org/10.1103/PhysRevB.39.8096} {\bibfield  {journal} {\bibinfo
  {journal} {Phys. Rev. B}\ }\textbf {\bibinfo {volume} {39}},\ \bibinfo
  {pages} {8096--8106} (\bibinfo {year} {1989})}\BibitemShut {NoStop}%
\bibitem [{\citenamefont {Mutlu}(1995)}]{mutlu1995}%
  \BibitemOpen
  \bibfield  {author} {\bibinfo {author} {\bibfnamefont {R.~H.}\ \bibnamefont
  {Mutlu}},\ }\bibfield  {title} {\enquote {\bibinfo {title} {Effect of
  nonlocal corrections to the local-density approximation on total-energy
  calculations of {K, Rb, and Cs}},}\ }\href
  {https://doi.org/10.1103/PhysRevB.52.1441} {\bibfield  {journal} {\bibinfo
  {journal} {Phys. Rev. B}\ }\textbf {\bibinfo {volume} {52}},\ \bibinfo
  {pages} {1441--1443} (\bibinfo {year} {1995})}\BibitemShut {NoStop}%
\bibitem [{\citenamefont {Kohn}\ and\ \citenamefont {Sham}(1965)}]{kohn1965}%
  \BibitemOpen
  \bibfield  {author} {\bibinfo {author} {\bibfnamefont {W.}~\bibnamefont
  {Kohn}}\ and\ \bibinfo {author} {\bibfnamefont {L.~J.}\ \bibnamefont
  {Sham}},\ }\bibfield  {title} {\enquote {\bibinfo {title} {Self-consistent
  equations including exchange and correlation effects},}\ }\href
  {https://doi.org/10.1103/PhysRev.140.A1133} {\bibfield  {journal} {\bibinfo
  {journal} {Phys. Rev.}\ }\textbf {\bibinfo {volume} {140}},\ \bibinfo {pages}
  {A1133--A1138} (\bibinfo {year} {1965})}\BibitemShut {NoStop}%
\bibitem [{\citenamefont {Ziesche}\ \emph {et~al.}(1998)\citenamefont
  {Ziesche}, \citenamefont {Kurth},\ and\ \citenamefont
  {Perdew}}]{ziesche1998}%
  \BibitemOpen
  \bibfield  {author} {\bibinfo {author} {\bibfnamefont {P.}~\bibnamefont
  {Ziesche}}, \bibinfo {author} {\bibfnamefont {S.}~\bibnamefont {Kurth}}, \
  and\ \bibinfo {author} {\bibfnamefont {J.~P.}\ \bibnamefont {Perdew}},\
  }\bibfield  {title} {\enquote {\bibinfo {title} {Density functionals from
  {LDA} to {GGA}},}\ }\href {https://doi.org/10.1016/S0927-0256(97)00206-1}
  {\bibfield  {journal} {\bibinfo  {journal} {Computational Materials Science}\
  }\textbf {\bibinfo {volume} {11}},\ \bibinfo {pages} {122--127} (\bibinfo
  {year} {1998})}\BibitemShut {NoStop}%
\end{thebibliography}%

\end{document}